\documentclass[hidelinks,12pt]{article}
\usepackage{amsmath}
\usepackage{placeins}
\usepackage{cancel}
\usepackage{bibunits}
\usepackage[normalem]{ulem}
\usepackage{relsize}
\allowdisplaybreaks
\usepackage{graphicx,psfrag,epsf,float}
\usepackage{enumerate}
\usepackage{enumitem}
\usepackage{adjustbox}
\usepackage{algorithm}
\usepackage{rotating} 
\usepackage{algpseudocode}
\usepackage{amssymb}
\usepackage{natbib}
\usepackage{geometry}
\usepackage{subfiles}
\usepackage{url}
\usepackage{amssymb,amsmath,amsfonts,eurosym,geometry,ulem,graphicx,caption,color,setspace,sectsty,comment,footmisc,caption,natbib,pdflscape,subfigure,array}
\usepackage{amsmath}
\usepackage{hyperref}  
\usepackage{booktabs} 
\newcommand{\blind}{0}

\newcommand{\bb}{\boldsymbol}
\newcommand{\expit}{\text{expit}}
\newcommand{\appropto}{\mathrel{\vcenter{
  \offinterlineskip\halign{\hfil$##$\cr
    \propto\cr\noalign{\kern2pt}\sim\cr\noalign{\kern-2pt}}}}}
\newgeometry{bottom=1in}
\newgeometry{left=1in,right=1in}
\usepackage{amsthm}
\newtheorem{theorem}{Theorem}
\newtheorem*{theorem*}{Theorem}
\newtheorem{lemma}{Lemma}
\newtheorem*{lemma*}{Lemma}

\newtheorem*{assumption*}{Assumption}
\newcommand{\beginsupplement}{%
        \renewcommand{\thesection}{S\arabic{section}}
\renewcommand{\thesubsection}{S2.\arabic{subsection}}
\renewcommand{\thelemma}{2}
}
 
\defaultbibliography{main}
\defaultbibliographystyle{apalike}
\title{\bf Heterogeneous Transfer Learning for Building High-Dimensional Generalized Linear Models with Disparate Datasets }

\author{
    Ruzhang Zhao \textsuperscript{1}  
    \and
    Prosenjit Kundu \textsuperscript{1,2}  
    \and
    Arkajyoti Saha \textsuperscript{1,3}  
    \and
    Nilanjan Chatterjee \textsuperscript{1,4}  \thanks{
     \texttt{nchatte2@jhu.edu}}
}

\begin{document}
\date{}
\maketitle
\footnotetext[1]{Department of Biostatistics,
    Johns Hopkins Bloomberg School of Public Health,
    Baltimore, MD, USA.}
\footnotetext[2]{Pfizer Inc.,
    Cambridge, MA, USA.}
\footnotetext[3]{Department of Statistics,
    University of Washington,
    Seattle, WA, USA.}
\footnotetext[4]{Department of Oncology, School of Medicine,
   Johns Hopkins University,
   Baltimore, MD, USA.}

\def\spacingset#1{\renewcommand{\baselinestretch}%
{#1}\scriptsize\normalsize} \spacingset{1.5}

%%%%%%%%%%%%%%%%%%%%%%%%%%%%%%%%%%%%%%%%%%%%%%%%%%%%%%%%%%%%%%%%%%%%%%%%%%%%%%

\iffalse
\if0\blind
{
  \bigskip
  \bigskip
  \bigskip
  \begin{center}
    {\LARGE\bf Title}
\end{center}
  \medskip
} \fi
\fi
\bigskip

%\author{Ruzhang Zhao, Prosenjit Kundu, Nilanjan Chatterjee}

\newpage
\begin{abstract}
Development of comprehensive prediction models are often of great interest in many disciplines of science, but datasets with information on all desired features often have small sample sizes. We describe a transfer learning approach for building high-dimensional generalized linear models using data from a main study with detailed information on all predictors and an external, potentially much larger, study that has ascertained a more limited set of predictors. We propose using the external dataset to build a reduced model and then “transfer” the information on underlying parameters for the analysis of the main study through a set of calibration equations which can account for the study-specific effects of design variables. We then propose a penalized generalized method of moment framework for inference and a one-step estimation method that could be implemented using standard \texttt{glmnet} package. We develop asymptotic theory and conduct extensive simulation studies to investigate both predictive performance and post-selection inference properties of the proposed method. Finally, we illustrate an application of the proposed method for the development of risk models for five common diseases using the UK Biobank study, combining information on low-dimensional risk factors and high throughout proteomic biomarkers.

%The development of comprehensive prediction models is often of great interest in many disciplines of science, but datasets with information on all desired features often have small sample sizes. In this article, we describe a transfer learning approach for building high-dimensional generalized linear models using data from a main study with detailed information on all predictors and an external, potentially much larger, study that has ascertained a more limited set of predictors. We propose using the external dataset to build a reduced model and then ``transfer" the information on underlying parameters for the analysis of the main study through a set of calibration equations while accounting for the study-specific effects of certain design variables. We then propose a penalized generalized method of moment framework for inference and a one-step estimation method that could be implemented using standard \texttt{glmnet} package after data augmentation and transformation. We develop asymptotic theory and conduct extensive simulation studies to investigate both predictive performance and post-selection inference properties of the proposed method. Finally, we illustrate an application of the proposed method for the development of risk prediction models for five common diseases using the UK Biobank study, combining baseline information from all study participants and recently released high-throughout proteomic data (\# protein = 1500) on a subset of the participants. 
\end{abstract}

%\noindent%

%We call our target individual dataset {\it internal}, while the source dataset {\it external} which we try to bring information to do data integration. The external data can be used in the format of individual data or summary statistics. 
\newpage
\begin{bibunit}
\section{Introduction}
As new technologies continue to expand our ability to measure high-dimensional features in many domains, there is growing interest in developing comprehensive models for predicting future outcomes, combining data from multiple domains, and using the algorithmic prowess of modern machine learning methods. However, in many applications, sufficiently large datasets that have ascertained information on features across all domains are hard to come across. Often, there are multiple datasets available, sometimes one nested in another, which include information on disparate but potentially overlapping sets of features. Specifically, across studies, there may be a small set of individuals who have information available on all of the features, but there would be large sets of individuals who will have partial information across the feature space.
\par 
In this article, we consider the setup of developing high-dimensional predictive models by combining data from a ``main study", which is assumed to have complete data on all of the features, and an ``external" study, which has information only on a partial set of features. In statistics, substantial literature exists for model fitting in low-dimensional settings incorporating external information. The range of methods includes, but is not limited to, various survey calibration techniques \citep{deville1992calibration,wu2001model,kim2010calibration}, empirical-likelihood \citep{qin1994empirical,chen1999pseudo,chaudhuri2008generalized,qin2009empirical},
%qin1999empirical,qin2000miscellanea,qin2015using,zhang2020generalized,sheng2021synthesizing,zhang2022integrative,fu2023integrative
constrained maximum-likelihood \citep{chatterjee2016constrained}, generalized method of moments \citep{imbens1994combining,kundu2019generalized} and ratio estimators \citep{rao1990estimating,taylor2023data}. We have earlier shown in a fairly general setting that the estimates of parameters for reduced-order models could be used to establish estimating functions informing parameters of a full model \citep{chatterjee2016constrained}, and information from a series of such functions can be concatenated through a generalized method of moment approach \citep{kundu2019generalized}.  Extensions of these methods for building high-dimensional models and downstream inference have largely remained unexplored.
\par 
In computer science, transfer learning (TL) is referred to as the use of a pre-trained model for a new but relevant task \citep{pan2010survey}. TL for building high-dimensional models incorporating %auxiliary 
information %from similar models 
from external studies has been considered in general settings \citep{li2022transfer,tian2022transfer}. TL methods have also been developed and tailored towards specific applications, such as for genetic risk scores \citep{zhao2022construction,tian2022multiethnic}, medical imaging \citep{raghu2019transfusion}, drug discovery \citep{cai2020transfer}, and reinforcement learning \citep{zhu2020transfer}. The problem we propose to tackle falls under heterogeneous transfer learning (HTL) \citep{day2017survey}, where external models are built based on lower-dimensional feature space than the target model. In general, HTL requires underlying ``translators'', analogous to calibration equations in survey settings, that allow information communication across different feature dimensions through projection into a shared space. While the formation of translators in specific applications of HTL has been described earlier, there has been no unified framework of HTL for building high-dimensional and interpretable regression models using multiple disparate datasets. 
\par

In this article, we build on our earlier work \citep{chatterjee2016constrained,kundu2019generalized} to propose a novel method termed HTL-GMM for building penalized regression models combining datasets with different features/covariates. The method assumes the availability of individual-level data on all of the desired features from a main study. Further, it is assumed that summary-level information on parameters associated with a ``reduced model'' is available from another, potentially larger study. We propose using the generalized method of moment (GMM) \citep{hansen1982large,imbens1996efficient} framework for unifying information available in different datasets into a joint objective function, accounting for study-specific effects of some key design variables. 
To deal with high dimensionality, we propose the regularization of regression parameters with standard penalization techniques and an additional regularization step for deriving the optimal weights for the underlying estimating functions. We then propose a one-step estimation technique starting with the initial estimator obtained from standard analysis of the main study alone and show that such a method could simply be implemented using the popular software package \texttt{glmnet} with suitable data augmentation and transformation. 
%We then propose the use of a penalized GMM approach \citep{caner2009Lasso} using either Lasso \citep{tibshirani1996regression} or adaptive Lasso \citep{zou2006adaptive,caner2014adaptive} for model regularization and develop scalable algorithms for implementing them using the powerful $\texttt{glmnet}$ package. 
We further derive asymptotic theory for the proposed method and consider post-selection asymptotic inference under the adaptive Lasso penalty function. %Regarding HTL, we do not make assumptions on the homogeneous population, while we have a weak plug-in estimator assumption~\ref{assumption:plugin}, which guarantees the parameters of reduced models from main and external studies are transferable. 
\par
The remainder of the article is organized as follows. In Section 2,  we describe the model setup and notations (Section 2.1), the penalized GMM framework (Section 2.2), the one-step estimation technique using $\texttt{glmnet}$ package (Section 2.3), and asymptotic theories (Section 2.4). In Section 3, we study the performance of the proposed method in terms of both predictive power and post-selection inference properties using simulated studies under linear and logistic regressions. In Section 4, we describe an application of the proposed method for the development of risk prediction models across several common complex diseases using data from the UK Biobank (UKB) study \citep{sudlow2015uk}. The article concludes with a discussion.

%Under HTL, the problem is referred as heterogeneous feature space or differing feature space \citep{day2017survey}, while in this work, we mainly focus on 

%Humans possess the ability to transfer knowledge across tasks and integrate information from diverse sources into a unified comprehension. This ability has inspired the concept of transfer learning, as detailed in \cite{torrey2010transfer}. Transfer learning adapts a pre-trained model to a new but relevant task, leveraging the knowledge gained from the original task \citep{weiss2016survey, pan2010survey}. The integration of disparate datasets is critical for improving predictive accuracy and facilitating robust statistical inference, including selective inference in high-dimensional scenarios. Transfer learning has been widely applied and shows its impressive performance in many fields, including medical imaging \citep{raghu2019transfusion}, drug discovery \citep{cai2020transfer}, reinforcement learning \citep{zhu2020transfer}, natural language processing \citep{houlsby2019parameter}, cross ancestry polygenic risk prediction \citep{zhang2022novel,zhang2023ensemble,jin2023me}, single-cell genomics \citep{wang2019data}, protein localization \citep{mei2011gene}. 
\par 
\par 
%The unmatched feature issue, specifically regarding the construction of full models using internal individual data, with the aid of summary-level information (or trained models) of reduced models from large external data, is discussed in \cite{chatterjee2016constrained}. Furthermore, \cite{zhai2022data} expands this concept from homogeneous population settings to heterogeneous settings.
%However, calibration information for data integration under high-dimensional models is not discussed. To fill in the gap, we propose heterogeneous transfer learning methods that account for unmatched feature spaces (akin to the settings outlined in \cite{chatterjee2016constrained}) within the framework of regularized high-dimensional Generalized Linear Models (GLMs). Similar like the UKB example, we have a main study (internal study) with limited size but details on all predictors. In contrast, the external study, despite its considerably larger size, only includes a subset of the predictors. The main study considers individual data, while the external study may utilize either individual data or summarized information. Here, we do not make assumptions on the homogeneous population, while we have a weak plug-in estimator assumption~\ref{assumption:plugin}, which guarantees the parameters of reduced models from main and external studies are transferable. 
\par 
\par 
%The rest of the article is organized as follows. Section~\ref{sec:pre}, the preliminary part, focuses on converting the unmatched feature space problem into notations, full model, and reduced model for GLMs. Section~\ref{sec:mainthm} discusses our main theorems and algorithm including efficient implementation. We demonstrate the numerical results, in particular, for prediction performance and selective inference in Section~\ref{sec:num}. Finally, we conclude with a discussion in Section~\ref{sec:con}. 

\section{Methods}
\label{sec:pre}

\subsection{Notations and Models setup}

%We aim to build risk prediction model on {\it main}/{\it internal} data with information transferred from {\it external} data.
%In a generalized linear model (GLM) setup, we denote the outcome of interest as $y$, the predictors as ${\bf x}$, and the corresponding coefficients as $\bb{\beta}$. 
We assume the target predictive model of interest for the  population underlying the main study takes the generalized linear model (GLM) \citep{mccullagh2019generalized} form,
\begin{equation}
y|{\bf x}\sim f_{\bb{\beta}}\left(y|{\bf x}\right) = \rho(y)\exp\left\{ y\cdot({\bf x}^\top\bb{\beta}) - \psi({\bf x}^\top\bb{\beta})\right\},
\label{eq:glm}
\end{equation}
%where $\bb{\beta}\in \mathbb{R}^{p_{\mathrm{X}}}$ ($p_{\mathrm{X}}$: feature size of ${\bf x}$).
where $y$ is the outcome variable, $\bf{x}$ is the vector of covariates/predictors of interest, and $\bb{\beta}$ is the vector of corresponding regression parameters. We assume ${\bf x}$ can be partitioned as ${\bf x}^\top = [{\bf a}^\top,{\bf z}^\top,{\bf w}^\top]$, where ${\bf a}$ denotes a set of study-specific confounding/adjustment factors, ${\bf z}$ denotes a set of variables available in both the main and external studies, and ${\bf w}$ denotes a set of variables uniquely observed in the main study. Correspondingly, we assume a partitioning of $\bb{\beta}$  as $\bb{\beta}^\top=[\bb{\beta}_{\mathrm{A}}^\top,\bb{\beta}_{\mathrm{Z}}^\top,\bb{\beta}_{\mathrm{W}}^\top]$. Below, we describe the proposed framework, assuming individual-level data are available from a main study and additional ``summary-level'' information is available from one external study. The framework can easily be extended to incorporate data from multiple external studies without much additional complexities. 
%and correspondingly, we assume a partition of the parameters $\bb{\beta}^\top = (\bb{\beta}^\top_{\mathrm{A}},\bb{\beta}^\top_{\mathrm{Z}},\bb{\beta}^\top_{\mathrm{W}})$ and the dimensions of both $\bb{\beta}_{\mathrm{Z}} $ and $\bb{\beta}_{\mathrm{W}} $ can be potentially high. $\rho$ and $\psi$ are known univariate functions. 
 %For linear regression, we can scale $y$, ${\bf x}$ to eliminate the intercept, while in general, intercept is included in study-specific adjustment factors for linear/logistic regression. 
We assume labelled data are available in the form $(y_i,{\bf x}_i)_{i=1}^{n}$ for $n$ independent individuals from the main study. %Correspondingly, we denote ${\bf y}$ and ${\bf X}$ the response vector and the matrix of predictors.
\par
%We assume capital letter ${\bf X}$ for matrix, while little letter ${\bf x}$ for vector. ${\bf x}_i$ denotes $i$th row of matrix ${\bf X}$; $y_i$ denotes $i$th element of vector ${\bf y}$. 
Throughout this article, we use superscript $\mathrm{E}$ to denote external study, and subscript $\mathrm{R}$ to denote covariate vector associated with reduced models. %and the expectation $\mathbb{E}$ is with respect to all randomness.
For the external study, we assume a predictive model for the response variable $y$ has been defined in terms of a vector of features $({{\bf x}^{\mathrm{E}}_{\mathrm{R}}})^\top = [\widetilde{\bf a}^\top,({\bf z}^{\mathrm{E}})^\top]$ %(corresponding matrix in external study, ${{\bf X}^{\mathrm{E}}_{\mathrm{R}}}= [\widetilde{\bf A},{{\bf Z}^{\mathrm{E}}}]$) 
, where $\widetilde{\bf a}$ denotes a set of study-specific variables, ${\bf z}^{\mathrm{E}}$ denotes the overlapping set of variables across the main and external studies. Specifically, we assume a ``reduced'' model of the form,
\begin{equation}
y|{\bf x}_{\mathrm{R}}\sim f_{\bb{\theta}}\left(y|{\bf x}_{\mathrm{R}}\right) = \rho\left({y}\right)\exp\left\{ 
    y\cdot ({\bf x}_{\mathrm{R}}^\top\bb{\theta})- \psi({\bf x}_{\mathrm{R}}^\top\bb{\theta})\right\},
\label{eq:glm:reduce}
\end{equation}
has been fitted to the external study, and underlying estimates of the parameters associated with the overlapping features ${\bf z}^{\mathrm{E}}$ and their uncertainties are available. We denote the sample size for the external study by $n^{\mathrm{E}}$.
%where $\bb{\theta}\in\mathbb{R}^{p_{\mathrm{X_{\mathrm{R}}}}}$ is the target coefficient for reduced model.
\par 
%We assume that if a reduced model (\ref{eq:glm:reduce}) is fitted to the main study, the true underlying regression parameters associated with the overlapping features ${\bf z}$ can be assumed to be the same as those from the external study. %, after the adjustment for respective study-specific variables, i.e., ${\bf a}$ or $\widetilde{\bf a}$. More formally, w
To connect the information from external study to the main study, we consider a reduced model of the form (\ref{eq:glm:reduce}) also for the main study and denote the partition of the reduced model parameters as $\bb{\theta}^\top = [\bb{\theta}_{\mathrm{A}}^\top,\bb{\theta}_{\mathrm{Z}}^\top]$, and $(\bb{\theta}^{\mathrm{E}})^\top = [\bb{\theta}_{\widetilde{\mathrm{A}}}^\top,(\bb{\theta}_{\mathrm{Z}}^{\mathrm{E}})^\top]$ for the main and external studies, respectively. Then, we assume that $\bb{\theta}_{\mathrm{Z}} = \bb{\theta}^{\mathrm{E}}_{\mathrm{Z}}$. Here, we note the importance of introducing the study-specific variables (${\bf a}$, $\widetilde{\bf a}$) and their effects  ($\bb{\theta}_{\mathrm{A}}, \bb{\theta}_{\widetilde{\mathrm{A}}})$ as they allow the assumption of ``transportability'' of reduced model parameters to hold only conditional on specific ``design'' variables. The incorporation of study-specific nuisance parameters for increasing the robustness of TL methods has been discussed earlier \citep{duan2022heterogeneity}. Our framework allows the incorporation of study-specific nuisance parameters in the HTL setting. In real applications, examples of such design variables may include intercept of logistic regression model to allow for different disease rates across populations, recruiting centers that are typically different across studies, and factors such as age, race, and sex, which may influence sampling/participation of subjects in different studies.
% For the reduced set of covariates ${\bf X}_{\mathrm{R}} = [{\bf A}, {\bf Z}]$, the target model takes the form
%In the external study, the missing of unmatched features ${\bf W}$ results in a different reduced model. We denote ${\bf X}_{\mathrm{R}} = [{\bf A},{\bf Z}]$ as the reduced version for the subset of explanatory variables. For external study, with different study-specific adjustment factors, we have ${\bf X}_{\mathrm{R}}^{\mathrm{E}} = \left[\widetilde{\bf A},{\bf Z}^{\mathrm{E}}\right]$. In general, we assume a reduced model with the subset of explanatory variables ${\bf X}_{\mathrm{R}}$ of the form,
\begin{assumption*}[A0, Transportability]
%We assume the reduced model underlying parameters of  overlapping variables, i.e. ${\bf z}$, between main and external studies are identical after adjusting for respective study-specific factors, i.e. $\bb{\theta}_{\mathrm{Z}}=\bb{\theta}_{\mathrm{Z}}^{\mathrm{E}}$. 
Conditional on the design variables $\mathrm{({\bf a}}$ and $\mathrm{\tilde{{\bf a}})}$, the regression parameters of the reduced model (\ref{eq:glm:reduce}) associated with overlapping features $\mathrm{({\bf z})}$ are transportable across the main and external studies. 
%We assume reduced model regression parameters associated with the overlapping features $\mathrm{({\bf z})}$, after accounting for study-specific adjustment factors, are transportable across populations underlying  the main and external studies.  
\label{sec:2.1}
\end{assumption*}
In the discussion section, we will further discuss strategies for relaxing the transportability assumption using shrinkage  estimation techniques. 
\subsection{Penalized GMM Framework}
In the absence of any external data, inference on the target model parameters $\bb{\beta}$ can be made by standard M-estimation theory based on estimating functions of the form
\begin{equation}
{\bf U}_{1,n}(\bb{\beta}) = \frac{1}{n}\sum_{i=1}^{n} {\bf U}_1({\bf x}_i,y_i;\bb{\beta})= \frac{1}{n}\sum_{i=1}^{n} \{
\mu({\bf x}_i^\top\bb{\beta})-y_i\} {\bf x}_i,
\label{eq:u1}
\end{equation}
%from population-level estimating equation $\mathbb{E}[ (\mu({\bf x}^\top\bb{\beta})-y) {\bf x}] = 0$. 
where  $\mu(s)=\partial{\psi(s)}/\partial{s}$. When $\bb{\beta}$ is high-dimensional, additional terms need to be incorporated for model regularization under different penalty functions. 
%Without external study, we can have good estimation for the $\bb{\beta}$ from Equation~\ref{eq:u1}. When $\bb{\beta}$ is high-dimensional, we can further add penalty terms in Equation~\ref{eq:u1}. 
\par
We observe that the reduced model parameters $\bb{\theta}$ for the main study are expected to satisfy the population estimating functions  $\mathbb{E}[\{\mu({\bf x}_{\mathrm{R}}^\top\bb{\theta}) - y\}{\bf z}] = 0$. If we rewrite the expectation in terms of the target outcome model $f_{\bb{\beta}}(y|{\bf x})$ (see \cite{chatterjee2016constrained}), we can show that $\mathbb{E}[\{\mu({\bf x}^\top\bb{\beta})-\mu({\bf x}_{\mathrm{R}}^\top\bb{\theta})\}{\bf z}] = 0$. This motivates us to consider the sample-level estimating functions of the form 
\begin{equation}
{\bf U}_{2,n}(\bb{\beta},\tilde{\bb{\theta}}) = \frac{1}{n}\sum_{i=1}^n {\bf U}_{2}({\bf x}_i;\bb{\beta},\tilde{\bb{\theta}}) = \frac{1}{n}\sum_{i=1}^n\{ \mu( {\bf x}_i^\top\bb{\beta}) - \mu( {\bf x}_{\mathrm{R},i}^\top\tilde{\bb{\theta}} )\}{\bf z}_{i},
\label{eq:u2}
\end{equation}
where $\tilde{\bb{\theta}}^\top = [ \hat{\bb{\theta}}_{\mathrm{A}}^\top, (\tilde{\bb{\theta}}_{\mathrm{Z}}^{\mathrm{E}})^\top]$, with $\hat{\bb{\theta}}_{\mathrm{A}}$ denoting the estimated effects of design variables ${\bf a}$ from the main study itself and $\tilde{\bb{\theta}}_{\mathrm{Z}}^{\mathrm{E}}$ denoting the estimated effects of the variables ${\bf z}$ from the external study. %After adjusting for ${\bf a}$ by $\hat{\bb{\theta}}_{\mathrm{A}}$ from main study, we can substitute the effect size of ${\bf z}$ in main study with its effect size from external study (plug-in estimator assumption~A0). 
%We propose estimating $\hat{\bb{\theta}}_{\mathrm{A}}$ from the main study by standard score functions associated with reduced model while fixing the the value of $\hat{\bb{\theta}}_{\mathrm{Z}}$ at its estimated value from the external study. The underlying estimating function takes the form ${\bf U}_{} = \sum_{i=1}^{n}{\bf a}_{i}{y_i-\mu(\bb{\theta}_{A}^{T}A_{i}+\hat\theta_{Z}^TZ_i) }$, where the terms $\hat{\bb{\theta}}_{\mathrm{Z}}^\top z_i$ are treated as offset terms while solving the equations for $\theta_{A}$.
We propose estimating $\hat{\bb{\theta}}_{\mathrm{A}}$ from the main study by standard score functions of the form,
\begin{equation}
{\bf U}_{3,n}(\bb{\theta}) = \frac{1}{n}\sum_{i=1}^{n} {\bf U}_3({\bf x}_{\mathrm{R},i},y_i;\bb{\theta}) = \frac{1}{n}\sum_{i=1}^n\{\mu({\bf a}_i^\top{\bb{\theta}}_{\mathrm{A}}+{\bf z}_i^\top{\bb{\theta}}_{\mathrm{Z}}) - y_i\}[{\bf a}_i^\top,{\bf z}_i^\top]^\top.
    \label{eq:glm:u3}
\end{equation} 
%where {\color{red} ${\bf x}_{\mathrm{R}}^\top=[{\bf a}^\top,{\bf z}^\top]$ \sout{${\bf x}_{\mathrm{R}}^\top=({\bf a}^\top,{\bf z}^\top)$} }denotes the covariate vector associated with the reduced model for the main study.
%where the terms $\hat{\bb{\theta}}_{\mathrm{Z}}^\top z_i$ are treated as offset terms while solving the equations for $\theta_{\mathrm{A}}$.
%{\color{red} \sout{We observe that Equation~(\ref{eq:u2}) is the key calibration equation, or the ``translator'', that allows ``transferring" of information on parameters of external models to the main study in a principled way that allows preservation of interpretation of parameters of the target model.}}  {\color{red} We observe that Equation~(\ref{eq:u2}) is the key calibration equation, or the ``translator''. It allows information ``transfer" on parameters of external models to the main study in a principled way that preserves the interpretation of the target model's parameters.} 
We observe that Equation~(\ref{eq:u2}) is the critical calibration equation, or the ``translator'', that allows the ``transferring" of information on parameters of external models to the main study in a principled way that preserves the interpretation of parameters of the target model. In general, the reduced model does not necessarily need to follow the same form as the target model, and the reduced model $f_{\bb{\theta}}(y|{\bf x}_{\mathrm{R}})$ needs not be correctly specified for the validity of the final inference \citep{chatterjee2016constrained}.
\par 
%%In Equation~\ref{eq:u2}, we use information of parameters from external model, i.e. $\tilde{\bb{\theta}}_{\mathrm{Z}}^{\mathrm{E}}$, to inform the parameters of the target model, i.e. $\bb{\beta}$. The plug-in estimator assumption~A0 offers us a bridge to connect main and external studies for overlapping features after adjusting for study-specific factors. %Here, we use one toy example to illustrate the intuition of the plug-in estimator assumption. 
% In UK Biobank framework, suppose the main study provides information on proteomics for a subset of UK Biobank participants and external study consists of the remaining participants, for whom we have no proteomics information. Now, we consider the following, ${\bf a}$: intercept; ${\bf z}$: age, gender; ${\bf w}$: proteomics; $y$: disease status. The disease case control ratio can be different in main and external studies, so we include the intercept in ${\bf a}$ for study-specific adjustment factors. The effect sizes for age and gender can be considered consistent across these two studies. In particular, in this example, the proteomics data is available for a subset of participants, it is reasonable to assume age and gender have the same effect sizes after adjusting for study-specific factors. 
%%Furthermore, we also discuss how to relax the assumption in our discussion section. 
\par 
We combine the estimating functions for $\bb{\beta}$ and define ${\bf U}_n(\bb{\beta},\tilde{\bb{\theta}})  = [{\bf U}_{1,n}(\bb{\beta})^\top$,${\bf U}_{2,n}(\bb{\beta},\tilde{\bb{\theta}})^\top]^\top=\frac{1}{n}\sum_{i=1}^n {\bf U}({\bf x}_i,y_i;\bb{\beta},\tilde{\bb{\theta}})$.
%, where ${\bf U}_{1,n}$ (\ref{eq:u1}) is derived using data only from the main study and ${\bf U}_{2,n}$ (\ref{eq:u2}) is the calibration equation to transfer information from the external to the main study. 
Here, ${\bf U}_{1,n}\in\mathbb{R}^{p_{\mathrm{X}}}$, and ${\bf U}_{2,n}\in\mathbb{R}^{p_{\mathrm{Z}}}$, where $p_{\mathrm{X}}$ and $p_{\mathrm{Z}}$ denote the dimensions of ${\bf x}$ and ${\bf z}$, respectively. Because the number of equations, $p_{\mathrm{X}}+p_{\mathrm{Z}}$, is larger than the number of parameters, $p_{\mathrm{X}}$, and $\bb{\beta}$ is potentially high-dimensional,  %Following the theory of Generalized Method of Moments (GMM) for estimation with over-identified equations and additional consideration of model regularization \citep{caner2009Lasso}
we propose a penalized GMM approach \citep{hansen1982large,imbens1996efficient,caner2009Lasso,caner2014adaptive} for inference through minimizing an objective function of the form, 
% =\frac{1}{2}n{\bf U}^\top{\bf CU}+ \textbf{P}_{\lambda,\alpha}(\bb{\beta})
\begin{equation}
Q(\bb{\beta},\tilde{\bb{\theta}}) = \frac{1}{2}n{\bf U}_n(\boldsymbol{\beta},\tilde{\bb{\theta}})^\top {\bf C}_n{\bf U}_n(\boldsymbol{\beta},\tilde{\bb{\theta}}) + \textbf{P}_{\lambda_n}(\bb{\beta}),
\label{eq:utcu}
\end{equation}
%We call it heterogeneous transfer learning via GMM ({\it HTL-GMM}, for short),
where ${\bf C}_n$ is a weighting matrix, $\textbf{P}_{\lambda_n}(\bb{\beta})$ is a nonconcave penalty function, and $\lambda_n$ denotes the underlying tuning parameter. Heretofore, we will refer to our proposed method, Heterogeneous Transfer Learning using GMM (HTL-GMM). In general, the method can be implemented using a variety of penalty functions,  including Lasso, Ridge, Elastic Net, adaptive Lasso \citep{tibshirani1996regression,zou2005regularization,zou2006adaptive}. In this article, we will focus on applications involving Lasso and adaptive Lasso, but our software could be used to implement the methods under other penalties implemented in the \texttt{glmnet} package. 

%Further, due to the contribution of the external study, the amount of information, i.e., the effective sample sizes, that contribute to the estimation of parameters associated ${\bf z}$ and ${\bf w}$ can be widely different. Therefore, one can consider regularization using distinct tuning parameters ($\lambda_{1,n}$, $\lambda_{2,n}$) for these two sets of variables. We describe this strategy as the bi-$\lambda$ strategy in simulation studies.
%Compared with ${\bf W}$, ${\bf Z}$ is paired with observation from external study ${\bf Z}^{\mathrm{E}}$, i.e. the sample size for ${\bf z}$ will be different from sample size for ${\bf w}$. We can adaptively choose different penalty levels for variables with different sample sizes. Other than regularization parameter $\lambda$, 
%We introduce ratio parameter $\alpha$ to assign different penalties, which is called bi-$\lambda$ strategy. 
\par 
%In GMM, the overall efficiency of estimator depends on estimation of ${\bf C}$, and the optimal $\widehat{\bf C}_{\mathrm{opt}}$ gives the most efficient estimator. Ignoring the penalty term, following Theorem 1 from \cite{kundu2019generalized}, the optimal $\widehat{\bf C}_{\mathrm{opt}}$ is given by $\mathrm{\bf V}_{\mathrm{opt}}^{-1}$, where $\sqrt{n}{\bf U}(\bb{\beta}^\star,\tilde{\bb{\theta}})\rightarrow \mathcal{N}(0,\mathrm{\bf V}_{\mathrm{opt}})$, with $n\rightarrow\infty$ and $\lim_{n\rightarrow +\infty} n^{\mathrm{E}}/n = r$, for some $0<r<\infty$, where $n^{\mathrm{E}}$ denotes the sample size of the external study. 

\subsection{One-step GMM Estimation }
\label{sec:algopart}
In this section, we develop a scalable implementation of the HTL-GMM algorithm utilizing the well-optimized package $\texttt{glmnet}$. We denote the vector ${\bf y} = [y_1,\cdots,y_n]^\top$, and the matrices ${\bf X} = [{\bf x}_1,\cdots,{\bf x}_n]^\top$, ${\bf Z} = [{\bf z}_1,\cdots,{\bf z}_n]^\top$, ${\bf X}_{\mathrm{R}} = [{\bf x}_{\mathrm{R},1},\cdots,{\bf x}_{\mathrm{R},n}]^\top$, and $\mu({\bf X}\bb{\beta}) = [\mu({\bf x}_1^\top\bb{\beta}),\cdots,\mu({\bf x}_n^\top\bb{\beta})]^\top$, $\mu({\bf X}_{\mathrm{R}}\bb{\theta}) = [\mu({\bf x}_{\mathrm{R},1}^\top\bb{\theta}),\cdots,\mu({\bf x}_{\mathrm{R},n}^\top\bb{\theta})]^\top$. We further assume ${\bf C}_n$ is positive definite and hence decomposable as ${\bf C}_n={\bf C}_n^{1/2}{\bf C}_n^{1/2}$, and $\hat{\bb{\beta}}_{n}^{(0)}$ is an initial ``well-behaved'' estimator of $\bb{\beta}$, which can be obtained using standard analysis of the main study. The GMM objective function (\ref{eq:utcu}), excluding the penalty term, can be approximated as a quadratic form and ignoring constant term as

\begin{equation}
\begin{aligned}
&{\bf U}_n(\boldsymbol{\beta},\tilde{\bb{\theta}})^\top {\bf C}_n{\bf U}_n(\boldsymbol{\beta},\tilde{\bb{\theta}})\\
&\approx \bb{\beta}^\top\left\{\frac{1}{n}{\bf X}_{\mathrm{ps}}(\hat{\bb{\beta}}_{n}^{(0)},\tilde{\bb{\theta}})^\top{\bf X}_{\mathrm{ps}}(\hat{\bb{\beta}}_{n}^{(0)},\tilde{\bb{\theta}})\right\}\bb{\beta} - 2\bb{\beta}^\top\left\{\frac{1}{n}{\bf X}_{\mathrm{ps}}(\hat{\bb{\beta}}_{n}^{(0)},\tilde{\bb{\theta}})^\top{\bf y}_{\mathrm{ps}}(\hat{\bb{\beta}}_{n}^{(0)},\tilde{\bb{\theta}})\right\},
\end{aligned}
\label{eq:ps}
\end{equation}
%+o_p(1)+ \mathrm{const.},
where a ``pseudo'' design matrix is defined as ${\bf X}_{\mathrm{ps}} (\hat{\bb{\beta}}_{n}^{(0)},\tilde{\bb{\theta}})= \sqrt{n}{\bf C}_n^{1/2}\frac{\partial {\bf U}_n(\hat{\bb{\beta}}_{n}^{(0)},\tilde{\bb{\theta}})}{\partial \bb{\beta}}$ and a ``pseudo'' response vector is defined as ${\bf y}_{\mathrm{ps}}(\hat{\bb{\beta}}_{n}^{(0)},\tilde{\bb{\theta}}) = \sqrt{n}{\bf C}_n^{1/2}\left\{\frac{\partial {\bf U}_n(\hat{\bb{\beta}}_{n}^{(0)},\tilde{\bb{\theta}})}{\partial \bb{\beta}}\hat{\bb{\beta}}_{n}^{(0)} - {\bf U}_n(\hat{\bb{\beta}}_{n}^{(0)},\tilde{\bb{\theta}})\right\}$. %respectively. Thus, for fixed values of tuning parameter $\lambda_n$, the penalized GMM solution can be potentially obtained by iterative calls to the powerful \texttt{glmnet} package. %We can compute ${\bf X}_{\mathrm{ps}}$ and ${\bf y}_{\mathrm{ps}}$ since ${\bf C}$ is always positive definite. $\mathrm{const.}$ is with respect to the target coefficient $\bb{\beta}$ with some fixed $\bb{\beta}^\star$ and $\tilde{\bb{\theta}}$. Classically, we optimize (\ref{eq:ps}) iteratively, where for any fixed ${\bf C}$, replace $\bb{\beta}^\star$ with an inital estimator of $\bb{\beta}$, compute ${\bf X}_{\mathrm{ps}}$ and ${\bf y}_{\mathrm{ps}}$ and refine the estimation of $\bb{\beta}$ until convergence. 
For linear regression, the derivation is exact with

%reminder term is exactly zero and the optimization of Equation~(\ref{eq:ps}) can be reduced to conducting least-square regression with
%the least-square form (\ref{eq:ps}) excluding the reminder term $o_p$, is the exact format without approximation. For ${\bf X}_{\mathrm{ps}}$, ${\bf y}_{\mathrm{ps}}$, they do not depend on $\bb{\beta}^\star$, and can be reduced to the following formats, and no iterative call is needed.
\begin{equation}
{\bf X}_{\mathrm{ps}} = \frac{1}{\sqrt{n}}{\bf C}_n^{1/2}[{\bf X},{\bf Z}]^\top{\bf X},\,\,\,\,\,\,\, {\bf y}_{\mathrm{ps}}(\tilde{\bb{\theta}}) =
\frac{1}{\sqrt{n}}{\bf C}_n^{1/2}\left[{\bf y}^\top{\bf X},  ({\bf X}_{\mathrm{R}}\tilde{\bb{\theta}})^\top{\bf Z}\right]^\top.
\label{eq:ps:linear}
\end{equation}
%Therefore, iterative computation will be degenerated to one-time computation. starting from ${\bf X}_{\mathrm{ps}}$ and ${\bf y}_{\mathrm{ps}}$, we can directly compute the final estimation of $\bb{\beta}$. 
For nonlinear regression, (\ref{eq:ps}) is derived by considering a second-order Taylor expansion and ignoring the term of the form 
${\frac{\partial^2 {\bf U}_n(\hat{\bb{\beta}}_n^{(0)},\tilde{\bb{\theta}})^\top}{\partial \bb{\beta}^\top\partial \boldsymbol{\beta}}}{\bf C}_n{\bf U}_n(\hat{\bb{\beta}}_n^{(0)},\tilde{\bb{\theta}})$, which is of the order $o_{p}(1)$ under stated regularity conditions (see Appendix). Under logistic regression, the pseudo design matrix and outcome take the form

%we observe ${\bf U}_n^\top{\bf C}_n{\bf U}_n$ is not convex but if we take the second order Taylor expansion of the GMM objective function with respect to $\bb{\beta}$ and ignore terms that are smaller order in sample sizes (See Lemma~\ref{lemma:2} in supplementary material), the problem can again be converted as a least-square problem. 
%but asymptotic convex. We take the second order Taylor expansion with respect to $\bb{\beta}$ around $\bb{\beta}^\star$ for quadratic approximation, where the form (\ref{eq:ps}) is asymptotic convex because the reminder term will asymptotically vanish. For ${\bf X}_{\mathrm{ps}}$ and ${\bf y}_{\mathrm{ps}}$, they depend on $\bb{\beta}^\star$ and have the following formats,
\begin{equation}
\begin{aligned}
    &{\bf X}_{\mathrm{ps}}(\hat{\bb{\beta}}_{n}^{(0)}) = \frac{1}{\sqrt{n}}{\bf C}_n^{1/2}[{\bf X},{\bf Z}]^\top{\bf D}(\hat{\bb{\beta}}_{n}^{(0)}){\bf X},\\
    &{\bf y}_{\mathrm{ps}}(\hat{\bb{\beta}}_{n}^{(0)},\tilde{\bb{\theta}}) = \frac{1}{\sqrt{n}}{\bf C}_n^{1/2}\left\{ [{\bf X},{\bf Z}]^\top\{{\bf D}(\hat{\bb{\beta}}_{n}^{(0)}){\bf X}   \hat{\bb{\beta}}_{n}^{(0)} -\expit({\bf X}\hat{\bb{\beta}}_{n}^{(0)})\}+[
     {\bf y}^\top{\bf X},  \expit({\bf X}_{\mathrm{R}}\tilde{\bb{\theta}})^\top {\bf Z}]^\top\right\},
\end{aligned}
\label{eq:ps:logistic}
\end{equation}
where ${\bf D}(\hat{\bb{\beta}}_{n}^{(0)}) = \mathrm{diag}\{\mathrm{dexpit}({\bf X}\hat{\bb{\beta}}_{n}^{(0)})  \}$, and $\mathrm{dexpit}(s)=\mathrm{expit}(s)\{1-\mathrm{expit}(s)\}$. 

The least-square form of Equation~(\ref{eq:ps}) immediately suggests that a solution to penalized GMM can be obtained based on the powerful $\texttt{glmnet}$ package with data augmentation and transformation. In general, one can iteratively compute ${\bf X}_{\mathrm{ps}}(\boldsymbol{\beta})$ and ${\bf y}_{\mathrm{ps}}(\boldsymbol{\beta},\tilde{\bb{\theta}})$, but we propose saving computation by considering only a one-step update starting from the initial estimator $\hat{\bb{\beta}}_{n}^{(0)}$, borrowing idea from one-step maximum likelihood estimation \citep{bickel1975one,van2006targeted, zheng2010asymptotic}. Our setting is particularly suitable for one-step estimation as an initial well-behaved (see regularity conditions) estimators $\hat{\bb{\beta}}_{n}^{(0)}$ is readily available based on the standard analysis of the main study. In our applications, we use standard Lasso analysis of the main study to derive the initial estimators, but many alternatives are possible. 
An overall algorithmic structure for our implementation of HTL-GMM is described in the appendix.

%In real  application, we need an initial estimate of $\bb{\beta}$. For example, we can first fit ridge regression with main study to get a good initial estimate of $\bb{\beta}$, then compute ${\bf X}_{\mathrm{ps}}(\bb{\beta}^\star,\tilde{\bb{\theta}})$, ${\bf y}_{\mathrm{ps}}(\bb{\beta}^\star,\tilde{\bb{\theta}})$ by (\ref{eq:ps}) and use them to estimate $\bb{\beta}$. In practice, the approximation is a way to avoid iterative optimization with compromising the efficiency of the estimator.
%Although this is not optimal solution since we do not iteratively get it, we can have estimation with great efficiency and simple computation by one step of iteration.
\par 

\subsection{Asymptotic Theory and Deriving Optimal Weight Matrix}
It is known that valid GMM inference  is possible for any choice of ${\bf C}_n$ as long as ${\bf C}_n \xrightarrow{p} {\bf C}$, where ${\bf C}$ is a finite positive definite and symmetric matrix. However, the 
%For GMM estimation, it is known that any choice of the weighting matrix ${\bf C}$ gives valid estimation, but the 
optimal choice of ${\bf C}_n$ is given by the inverse of the asymptotic variance-covariance matrix for the underlying estimating functions ${\bf U}_n(\bb{\beta}^\star,\tilde{\bb{\theta}})$. %, assuming $\sqrt{n}$ consistency and asymptotic normality, i.e., ${\bf C}_{\mathrm{opt}} = \mathrm{\bf V}_{\mathrm{opt}}^{-1}$. 
In the following lemma, we describe the asymptotic variance-covariance matrix of ${\bf U}_n(\bb{\beta}^\star,\tilde{\bb{\theta}})$, taking into account uncertainty associated with $\tilde{\bb{\theta}}$.
\par
Let ${\bf V}_{{\bf x},{\bf x}} =\mathbb{E}\left[{\bf x}{\bf x}^\top\{\mu({\bf x}^\top\bb{\beta}^\star) - y\}^2\right]$, ${\bf V}_{{\bf z},{\bf z}} = \mathbb{E}\left[{\bf z}{\bf z}^\top\{\mu({\bf x}^\top\bb{\beta}^\star) - \mu({\bf x}_{\mathrm{R}}^\top\bb{\theta}^\star)  \}^2\right]$, ${\bf V}_{{\bf x},{\bf z}} =\mathbb{E}\left[{\bf x}{\bf z}^\top\{\mu({\bf x}^\top\bb{\beta}^\star) -  y\}\{\mu({\bf x}^\top\bb{\beta}^\star) - \mu({\bf x}_{\mathrm{R}}^\top\bb{\theta}^\star)\}\right]$, ${\bf V}_{{\bf x},{\bf x}_{\mathrm{R}}} = \mathbb{E}\left[{\bf x}{\bf x}_{\mathrm{R}}^\top\{\mu({\bf x}^\top\bb{\beta}^\star) - y\}\{\mu({\bf x}_{\mathrm{R}}^\top\bb{\theta}^\star) - y\}\right]$, and ${\bf V}_{{\bf z},{\bf x}_{\mathrm{R}}}  = \mathbb{E}\left[{\bf z}{\bf x}_{\mathrm{R}}^\top\{\mu({\bf x}^\top\bb{\beta}^\star) - \mu({\bf x}_{\mathrm{R}}^\top\bb{\theta}^\star)\}\{\mu({\bf x}_{\mathrm{R}}^\top\bb{\theta}^\star) - y\}\right]$, where throughout $\bb{\beta}^\star$ and $\bb{\theta}^\star$ indicate true values of the respective parameters in the underlying population for the main study. We denote $\bb{\Gamma}_{{\bf s},{\bf t}} = \mathbb{E}\left[{\bf st}^\top \mu^\prime({\bf x}_{\mathrm{R}}^\top\bb{\theta}^\star) \right] $, where ${\bf s}$ and ${\bf t}$ are subvectors of ${\bf x}$, and $\mu^\prime(\cdot)$ is the derivative function. Further define the partition $\bb{\Gamma}_{ {\bf x}_{\mathrm{R}},{\bf x}_{\mathrm{R}}}^{-1} = [\bb{\Gamma}^{ {\bf x}_{\mathrm{R}},{\bf a}},\bb{\Gamma}^{ {\bf x}_{\mathrm{R}},{\bf z}}]$. 
%{\color{red} for its first $p_{\mathrm{A}}$ and last $p_{\mathrm{Z}}$ columns, respectively.}
We assume asymptotic normality of parameter estimates from external study as
%of both $\hat{\bb{\theta}}_{\mathrm{A}}$ and $\tilde{\bb{\theta}}_{\mathrm{Z}}^{\mathrm{E}}$ so that $\sqrt{n}(\hat{\bb{\theta}}_{\mathrm{A}}-\bb{\theta}^\star_{\mathrm{A}}) \xrightarrow{d} \mathcal{N}({\bf 0},{\bf V}_{\bb{\theta}_{\mathrm{A}}})$ and
$\sqrt{n^{\mathrm{E}}}(\tilde{\bb{\theta}}_{\mathrm{Z}}^{\mathrm{E}}-\bb{\theta}^\star_{\mathrm{Z}}) \xrightarrow{d} \mathcal{N}({\bf 0},{\bf V}_{ \bb{\theta}_{\mathrm{Z}}^{\mathrm{E}} })$. 
\begin{lemma} 
\label{lemma:U}
Under the Assumptions (A0)-(A4) stated in the appendix and assuming that $\lim_{n\rightarrow +\infty} n^{\mathrm{E}}/n  = r, 0<r<\infty$,  $\sqrt{n}{\bf U}_n(\bb{\beta}^\star,\tilde{\bb{\theta}}) = \sqrt{n}{\bf U}_n(\bb{\beta}^\star,\hat{\bb{\theta}}_{\mathrm{A}},\tilde{\bb{\theta}}_{\mathrm{Z}}^{\mathrm{E}})\xrightarrow{d} \mathcal{N}({\bf 0},\mathrm{\bf V})$, where  %where $\mathrm{\bf V} = \mathrm{\bf V}(\bb{\beta}^\star,\bb{\theta}^\star,r)$ follows 
\begin{align*}
&\mathrm{\bf V}= \left(\begin{array}{ccc}
  {\bf V}_{11}   & {\bf V}_{12} \\
    {\bf V}_{12}^\top& {\bf V}_{22}
\end{array}\right), \text{ with } {\bf V}_{11}={\bf V}_{{\bf x},{\bf x}},\\
&{\bf V}_{12} = {\bf V}_{{\bf x},{\bf z}}+{\bf V}_{{\bf x},{\bf x}_{\mathrm{R}}}\bb{\Gamma}^{{\bf x}_{\mathrm{R}},{\bf a}}\bb{\Gamma}_{{\bf a},{\bf z}},\\
&{\bf V}_{22}={\bf V}_{{\bf z},{\bf z}} + \bb{\Gamma}_{{\bf z},{\bf x}_{\mathrm{R}}} \left(\begin{array}{cc}{\bf V}_{\bb{\theta}_{\mathrm{A}}}  & {\bf 0} \\
         {\bf 0}& r^{-1}{\bf V}_{\bb{\theta}_{\mathrm{Z}}^{\mathrm{E}}}
    \end{array}\right)
    \bb{\Gamma}_{{\bf z},{\bf x}_{\mathrm{R}}}^\top+{\bf V}_{{\bf z},{\bf x}_{\mathrm{R}}}\bb{\Gamma}^{{\bf x}_{\mathrm{R}},{\bf a}}\bb{\Gamma}_{{\bf a},{\bf z}}+({\bf V}_{{\bf z},{\bf x}_{\mathrm{R}}}\bb{\Gamma}^{{\bf x}_{\mathrm{R}},{\bf a}}\bb{\Gamma}_{{\bf a},{\bf z}})^\top.
\end{align*}
\end{lemma}

A sketch of the proof is provided in the appendix. We denote $ \widehat{\mathrm{\bf V}}_{\mathrm{opt},n}= \widehat{\mathrm{\bf V}}_{n}(\hat{\bb{\beta}}_n^{(0)},\tilde{\bb{\theta}},\hat{r})$ as the sample level estimator of $\mathrm{\bf V}=\mathrm{\bf V}(\bb{\beta}^\star,\bb{\theta}^\star,r)$ obtained by plugging in $\hat{\bb{\beta}}_{n}^{(0)}$, $\tilde{\bb{\theta}}$ and $\hat{r}$ and replacing all expectations in the formula with their empirical versions.
%One choice of ${\bf C}_n$ is $\widehat{\mathrm{\bf V}}_{\mathrm{opt},n}^{-1}(\hat{\bb{\beta}},\tilde{\bb{\theta}},\hat{r})$. 
%{\color{red}\sout{In practice, one can use a one-step estimator of ${\bf C}_{\mathrm{opt},n}$ where an initial estimate of $\bb{\beta}$ is obtained, e.g., the one that could be obtained using standard ridge regression with cross-validation on the data from main study only, and then plug in the value of the initial estimate in the derivation of ${\bf C}_{\mathrm{opt},n}$.} 
%For obtaining optimal performance of GMM, one can use ${\bf C}_{\mathrm{opt},n}=\widehat{\mathrm{\bf V}}_{\mathrm{opt},n}^{-1}(\hat{\bb{\beta}},\tilde{\bb{\theta}},\hat{r})$. 
%In practice, one can use a one-step estimator of ${\bf C}_{\mathrm{opt},n}$, where an initial estimate of $\bb{\beta}$ is obtained and plugged in the derivation of ${\bf C}_{\mathrm{opt},n}$. An example of the initial estimate of $\bb{\beta}$ is to use ridge regression with cross-validation on the main study data only. 
\par 
According to standard GMM theory \citep{hansen1982large}, the use of the weight matrix ${\bf C}_{\mathrm{opt},n}=\widehat{\mathrm{\bf V}}_{\mathrm{opt},n}^{-1}$ will lead to asymptotic efficiency. In high-dimensional settings, however, when $n/p$ is small,  the use of  ${\bf C}_{\mathrm{opt},n}$ may lead to poor performance of GMM. First, when $n \leq p_{\mathrm{X}}+p_{\mathrm{Z}}$, $\widehat{\bf V}_{n}(\hat{\bb{\beta}}_n^{(0)})$ would be ill-conditioned and not invertible. Even when $n>p_{\mathrm{X}}+p_{\mathrm{Z}}$ but $n$ is not adequately large,  $\widehat{\bf V}_{n}(\hat{\bb{\beta}}_n^{(0)})$ will be a noisy estimator of $\widehat{\bf V}_{n}(\bb{\beta})$. We note that regularization of the initial estimator of $\bb{\beta}$ itself is not adequate here, and a separate step is needed for regularization of the variance-covariance matrix. We propose the use of variational GMM (vGMM), initially developed in the context of non-parametric instrumental variable analysis \citep{bennett2023variational}, to consider an objective function of the form
\begin{equation*}
{\bf U}_n( \bb{\beta},\tilde{\bb{\theta}}  )^\top\left(\widehat{\mathrm{\bf V}}_{\mathrm{opt},n}+\alpha_n {\bf K}_n\right)^{-1}{\bf U}_n( \bb{\beta},\tilde{\bb{\theta}}),
\end{equation*}
%= \arg \min_{\bb{\beta}\in\mathbb{R}^{p_{\mathrm{X}}}} \sup_{{\bf s}\in\mathbb{R}^{p_{\mathrm{X}}+p_{\mathrm{Z}}}}  {\bf s}^\top {\bf U}_n( \bb{\beta},\tilde{\bb{\theta}} )  -\frac{1}{4}{\bf s}^\top\widehat{\mathrm{\bf V}}_{\mathrm{opt},n}{\bf s} - R_n({\bf s}) \\
where $\alpha_n$ is a tuning parameter, and ${\bf K}_n$ is a positive semi-definite kernel matrix. While all ${\bf K}_n$ will give rise to asymptotically efficient results as long as $\alpha_n=o(1)$, a good choice of ${\bf K}_n$ is important for the efficiency of the proposed method in a finite sample. We choose kernel function of the form ${\bf K}_n = \Big(\begin{array}{@{}c@{}@{}c@{}}
  {\bf K}_{11,n}\,\, & {\bf 0} \\[-1.5ex]
    {\bf 0}&{\bf 0}
\end{array}\Big)$, so that the regularization has an effect only on the variance-covariance of the high-dimensional component of the estimating function, i.e., ${\bf U}_{1,n}$.  The most obvious choice of ${\bf K}_{11,n}$ is $\mathbf{I}_n$ which corresponds to the ridge penalty. However, our extensive numerical exploration indicates that the choice of the optimal kernel function may depend on the task, e.g., prediction versus classification. We find that for classification under logistic regression, a superior choice for the kernel function is ${\bf K}_{11,n} = \widehat{\bf V}_{11}$ which essentially does a constant multiplicative shrinkage to the elements of ${\bf U}_{1,n}$ relative to those of ${\bf U}_{2,n}$. Interestingly, multiplicative shrinkage and tuning strategy has been considered in the past for knowledge distillation under multiple loss functions \citep{hinton2015distilling, yim2017gift} and has been shown to work well for classification tasks but not as much in regression tasks \citep{clark2019bam,takamoto2020efficient}.

The asymptotic property of GMM estimators with shrinkage penalties has been described in the past \citep{caner2009Lasso,caner2014adaptive}. In particular, it has been shown that the GMM estimator with adaptive Lasso enjoys oracle property. In our setting, additional complexities arise due to the use of plug-in estimate $\tilde{\bb{\theta}}$ in the GMM objective function and the uncertainty associated with it. In the following theorems, we use Lemma~\ref{lemma:U} and the property of the one-step estimator to establish that the well-known properties of adaptive Lasso are expected to hold for penalized GMM estimation in our setting. Let $\mathcal{A}_{\mathrm{Z}}$ and $\mathcal{A}_{\mathrm{W}}$ denote the index sets of subvectors of ${\bf x}$ associated with the overlapping (${\bf z}$) and unmatched (${\bf w}$) features and denote $\mathcal{A}^\star$ as the index set of a subset of variables of ${\bf x}$ that have true nonzero effects on the outcome under the target GLM model (\ref{eq:glm}). Now, let the penalty function associated with adaptive Lasso be defined as
%it performs as well as if the true underlying model given in advance, whereas GMM estimator with Lasso penalty only has asymptotic consistency. Here, we take the penalty term 
$\textbf{P}_{\lambda_n}(\bb{\beta}) = \lambda_{n}\sum_{j} \hat{w}_j|\beta_j|$. The weight vector $\hat{\bb{w}} = 1/|\hat{\boldsymbol{\beta}}^{\mathrm{glm},(0)}|^{{}^{\mbox{\scalebox{0.75}{$\gamma$}}} }$ or $\hat{\bb{w}} = 1/|\hat{\boldsymbol{\beta}}^{\mathrm{ridge},(0)}|^{{}^{\mbox{\scalebox{0.75}{$\gamma$}}} }$is derived using data from the main study by standard GLM or ridge regression \citep{zou2006adaptive}. Further, $\gamma$ is a scale factor assumed to be positive, often in practice, picked as 1/2, 1, or 2. 
%{\color{red}\sout{Let $\mathcal{A}^\star$ denote the index set of subset of variables of ${\bf x}$ that have true nonzero effects on the outcome under the target GLM model (\ref{eq:glm}). (This is moved upwards) }}
The following theorem states the asymptotic properties of the one-step HTL-GMM estimator under adaptive Lasso, denoted as $\hat{\boldsymbol{\beta}}^{\mathrm{aLasso},(n)}$. To proceed, we first denote $\bb{\Gamma}_{\bb{\beta}} = \mathbb{E}[\frac{\partial  {\bf U}({\bf x},y;\bb{\beta}^\star,\bb{\theta}^\star)}{\partial \bb{\beta}}]$ and use the subscript $\mathcal{A}^\star$ to indicate sub-matrices containing rows and columns associated with variables included in the index set $\mathcal{A}^\star$.
 %Denote $\hat{\boldsymbol{\beta}}^{\mathrm{aLasso}} = \arg\min_{\boldsymbol{\beta}}Q(\boldsymbol{\beta})$ from (\ref{eq:utcu}) for adaptive Lasso penalty.
\begin{theorem}~(HTL-GMM Adaptive Lasso Oracle Property) Under the Assumptions (A0)-(A6) stated in the appendix, and supposing that $\lambda_{n}/\sqrt{n}\rightarrow 0 $ and $\lambda_{n} n^{(\gamma - 1)/2}\rightarrow +\infty$ for $\gamma >0$, $\lim_{n\rightarrow \infty}n^{\mathrm{E}}/n = r, 0<r<\infty$, 
%Then, the HTL-GMM with adaptive Lasso estimator, $\hat{\boldsymbol{\beta}}^{\mathrm{aLasso},(n)}$ must satisfy the following properties, \\
the following results will hold for estimator $\hat{\boldsymbol{\beta}}^{\mathrm{aLasso},(n)}$: \\
1. Consistency in variable selection: $\lim_n\mathbb{P}\left(\mathcal{A}_n=\mathcal{A}^\star\right)=1$, where $\mathcal{A}_{n} = \{j:\hat{\beta}^{\mathrm{aLasso},(n)}_j \neq 0 \}$. \\
2. Asymptotic normality: $\sqrt{n}\left(  
\hat{\boldsymbol{\beta}}^{\mathrm{aLasso},(n)}_{\mathcal{A}^\star} - \bb{\beta}^\star_{\mathcal{A}^\star}\right) \xrightarrow{d} \mathcal{N}\left({\bf 0}, \boldsymbol{\Sigma}^\star_{\mathcal{A}^\star}  \right)$,
where \\$\boldsymbol{\Sigma}^\star_{\mathcal{A}^\star}= \{(\bb{\Gamma}_{\bb{\beta}}^\top{\bf C}\bb{\Gamma}_{\bb{\beta}})_{\mathcal{A}^\star}\}^{-1}\{( \bb{\Gamma}_{\bb{\beta}}^\top{\bf C}\mathrm{\bf V}{\bf C}\bb{\Gamma}_{\bb{\beta}} )_{\mathcal{A}^\star}\} \{ ( \bb{\Gamma}_{\bb{\beta}}^\top{\bf C}\bb{\Gamma}_{\bb{\beta}} )_{\mathcal{A}^\star}\}^{-1}$.\\
3. Optimality: The choice of ${\bf C}_n=\left(\widehat{\mathrm{\bf V}}_{\mathrm{opt},n}+\alpha_n{\bf K}_n\right)^{-1}=\left(\widehat{\mathrm{\bf V}}_{n}(\hat{\bb{\beta}}_n^{(0)},\tilde{\bb{\theta}},\hat{r})+\alpha_n{\bf K}_n\right)^{-1}$, $\alpha_n = o(1)$, leads to asymptotic optimality with $\bb{\Sigma}^\star_{\mathcal{A}^\star}= \{(\bb{\Gamma}_{\bb{\beta}}^\top\mathrm{{\bf V}^{-1}}\bb{\Gamma}_{\bb{\beta}})_{\mathcal{A}^\star}\}^{-1}$.
\label{thm:oracle2}
\end{theorem}
% \begin{proof} The sketch Intuitively, the proof has two main steps. (1) We use \cite{zou2006adaptive} to show that under the assumption of Theorem~\ref{thm:oracle2}, for our HTL-GMM estimation to enjoy oracle property, a sufficient condition is $\sqrt{n}{\bf U}(\bb{\beta}^\star,\tilde{\bb{\theta}})$ to be asymptotic normal. (2) Using Taylor expansion, we show that this is equivalent to asymptotic normality of $\sqrt{n}(\tilde{\bb{\theta}} - \bb{\theta}^\star)$, where $\bb{\theta}^\star$ is the population counterpart of $\tilde{\bb{\theta}}$. We recall that $\tilde{\bb{\theta}}$ may come from a high-dimensional model, which makes it a non-trivial problem. We use the tools in \cite{voorman2014inference} to establish the asymptotic normality of $\sqrt{n}(\tilde{\bb{\theta}}-\bb{\theta}^\star)$. Please see supplementary material for a detailed proof. \end{proof}

%\subsubsection{Post-Selection Inference}
%As stated in \cite{zou2006adaptive}, the Lasso selection could be inconsistent unless we sacrifice the rate of convergence, and the corresponding GMM estimator with Lasso penalty is stated consistent under some strict conditions \citep{caner2009Lasso}, while the GMM estimator with adaptive Lasso \citep{caner2014adaptive} is shown to enjoy oracle property just like adaptive Lasso. Therefore, we take adaptive Lasso in $\textbf{P}_{\lambda}(\bb{\beta})$ to perform the variable selection and post-selection inference. 
\par 

The proof and conditions for Theorem~\ref{thm:oracle2} are derived by extending those in \cite{zou2006adaptive} (see Appendix). Further, $\boldsymbol{\Sigma}^\star_{\mathcal{A}^\star}$ can be empirically estimated as 
$\left[\{{\bf X}_{\mathrm{ps}}(\hat{\bb{\beta}}_n)^\top{\bf X}_{\mathrm{ps}}(\hat{\bb{\beta}}_n)\}_{\mathcal{A}_n}\right]^{-1}$ $\left[\{{\bf X}_{\mathrm{ps}}(\hat{\bb{\beta}}_n)^\top{\bf C}_n^{1/2}\widehat{\mathrm{\bf V}}_{n}(\hat{\bb{\beta}}_n){\bf C}_n^{1/2}{\bf X}_{\mathrm{ps}}(\hat{\bb{\beta}}_n)\}_{\mathcal{A}_n}\right]$ $\left[\{{\bf X}_{\mathrm{ps}}(\hat{\bb{\beta}}_n)^\top{\bf X}_{\mathrm{ps}}(\hat{\bb{\beta}}_n)\}_{\mathcal{A}_n}\right]^{-1}$ where we denote $\hat{\bb{\beta}}_n$ as the final estimation from HTL-GMM.
%$\left\{\left({{\bf X}_{\mathrm{ps}}^{(0)}}^\top{\bf X}_{\mathrm{ps}}^{(0)}\right)_{\mathcal{A}_n}\right\}^{-1}$ $\left\{\left({{\bf X}_{\mathrm{ps}}^{(0)}}^\top{\bf C}_n^{1/2}\widehat{\bf V}_{\mathrm{opt},n}{\bf C}_n^{1/2}{\bf X}_{\mathrm{ps}}^{(0)}\right)_{\mathcal{A}_n}\right\}$ $\left\{\left({{\bf X}_{\mathrm{ps}}^{(0)}}^\top{\bf X}_{\mathrm{ps}}^{(0)}\right)_{\mathcal{A}_n}\right\}^{-1}$, where we denote ${\bf X}_{\mathrm{ps}}^{(0)} = {\bf X}_{\mathrm{ps}}(\hat{\bb{\beta}}_n^{(0)})$.
\section{Simulation Studies}
\label{sec:num}
\subsection{Prediction Performance}
\label{sec:predict}
We simulate data assuming all covariates (${\bf x}$) follow a multivariate normal distribution and allow for complex correlation structures within and between overlapping (${\bf z}$) and unmatched features (${\bf w}$). We consider $\mathrm{dim}({\bf z}) = p_{\mathrm{Z}}$ to be 10 or 40, while $\mathrm{dim}({\bf w}) = p_{\mathrm{W}}$ is chosen to be 150 or 1,500. Both sets of variables, i.e., ${\bf z}$ and ${\bf w}$, are partitioned into independent blocks of tens and assumed to have autocorrelation structures within blocks with the correlations among the nearest neighbors fixed at 0.5. We assume the number of non-null variables in ${\bf z}$ is 10, i.e., all of them are true predictors when $p_{\mathrm{Z}}=10$ and a quarter of them are true predictors when $p_{\mathrm{Z}}=40$, respectively. We assume the number of non-null variables in ${\bf w}$ is 15, i.e., only ten percent or one percent of them are true predictors, depending on $p_{\mathrm{W}}=150$ or $p_{\mathrm{W}}=1500$, respectively. When $p_{\mathrm{Z}}=40$, we randomly pick three variables from the first and third blocks and two from the second and fourth blocks to have non-null effects. When $p_{\mathrm{W}}=150$, we randomly pick one variable from each of the ten blocks to have non-null effects. When $p_{\mathrm{W}}=1500$,  we follow the same scheme to assign the non-null effects within the first ten blocks and then assume the remaining blocks include only noise variables. We further allow ten pairs of non-null variables across ${\bf z}$ and ${\bf w}$ to have correlation with a fix value of $\rho=0.3$.
%, {\color{red}e.g., the 1st non-null variable of ${\bf z}$ is correlated with the 13th non-null variable of ${\bf w}$.} %We also allow 10 pairs of correlation between 8 out of 10 non-null ${\bf z}$ and 10 out of 15 non-null ${\bf w}$ variables by linking variables from the two sets as pairs with correlation 0.3. 

%“We further allow 10 pairs of non-null variables across ${\bf z}$ and ${\bf w}$ to be in correlation with a value of 0.3. The following pairs from $({\bf z}, {\bf w}$ are selected randomly to be correlated: (1,11), (1,13), (2,14), (3,10), (4,4), (5,3), (6,2), (8,1), (10, 8), (10,9). For example, (1,11) means the 1st non-null variable of ${\bf z}$ is correlated with the 11th non-null variable of ${\bf w}$.”
\par
\begin{figure}[h!]
    \centering
    \includegraphics[width=.8\linewidth]{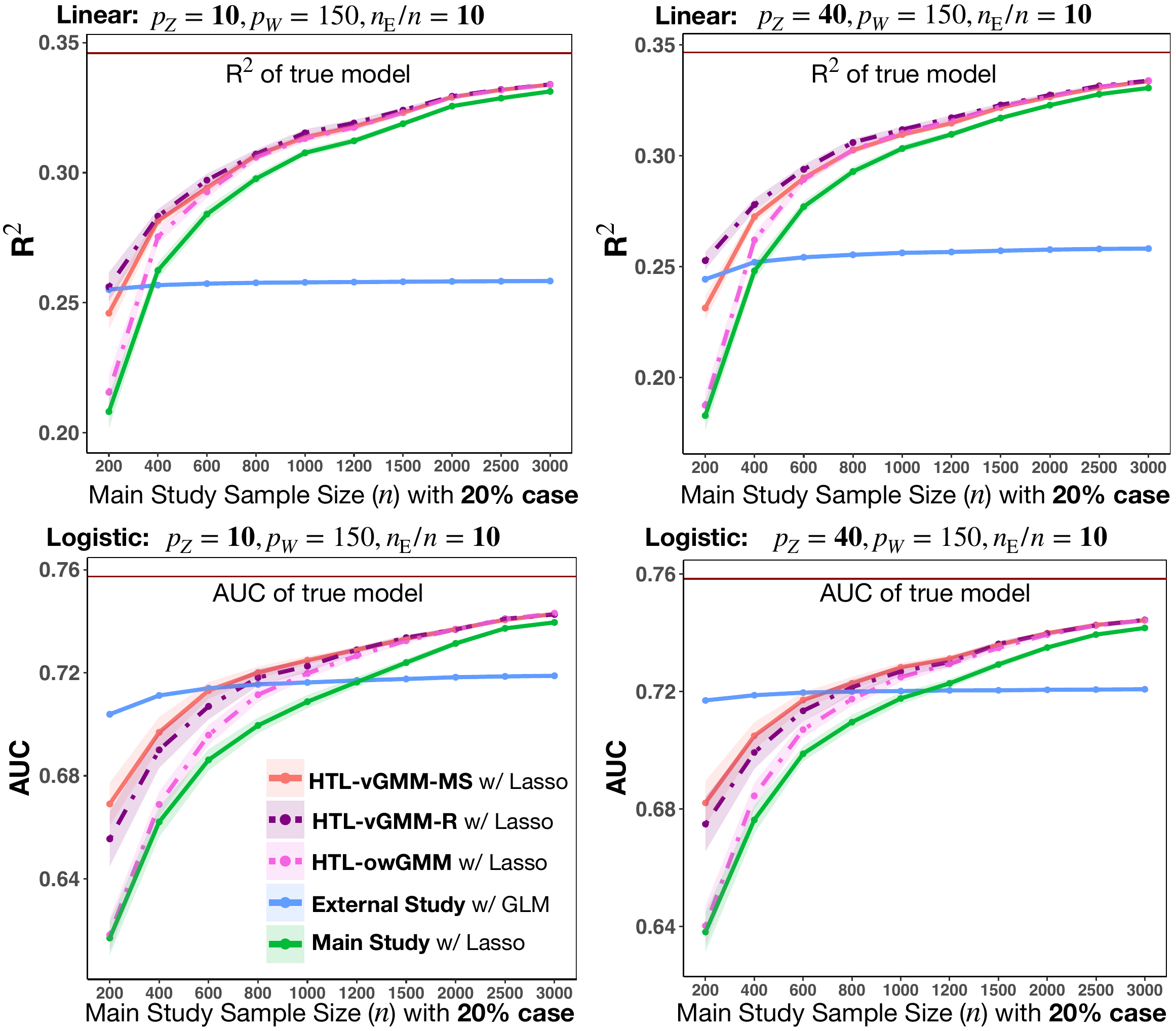}
    \caption{\textbf{Simulation study results showing the predictive performance of HTL-GMM and alternative methods with Lasso penalization for linear and logistic regression models.}
    The methods included are HTL-GMM-MS which uses multiplicative-shrinkage variational kernel,  HTL-GMM-R which uses ridge variational kernel, HTL-owGMM which uses ordinary GMM, standard analysis of only the main study  or external study in the linear or logistic regression setting. For HTL-GMM and standard analysis of the main study, the full models are fitted using linear or logistic regressions with the Lasso penalty function. The reduced model for the external study is fitted using standard linear or logistic regression. The prediction performance of models, quantified by $\mathrm{R}^2$ (first row) or AUC (last row), is evaluated based on a large validation sample simulated independent of the main and external studies. The sample size of the main study varies along the $x$-axis for each figure. The number of overlapping variables ($p_{\mathrm{Z}}$) varies across different panels while the sample size of the external study relative to the main study ($n^{\mathrm{E}}/n$) is 10.}
    \label{fig:lgLasso}
\end{figure}

 
Conditional on ${\bf X}$, we generate continuous and binary outcomes based on linear or logistic regression models, respectively. For linear regression, we assume ${\bf y}$ and the columns of ${\bf X}$ are centered to have mean zero, thus avoiding the need to estimate the intercept parameter. Further, the effect sizes of non-null variables and the residual variances are chosen so that $\mathrm{R}^2$ associated with the true model is 0.343. For the logistic regression model, the underlying intercept parameter is chosen to fix $\mathrm{Pr}(y=1)$ in the population to be 0.2. The effect sizes of the non-null variables are chosen so that the area under the ROC curve (AUC) statistics associated with the true model is 0.754. We simulate individual-level data on $(y,{\bf x})$ using the same model for the main and external studies. We assume the sample size for the external study is 10-fold that of the main study. For data analysis, we first fit a reduced linear or logistic regression model with only ${\bf Z}$ as the covariates using the external data and then only use the underlying estimates and variance-covariance matrices for subsequent HTL-GMM analysis. We use 10-fold cross-validation within the main study for each simulated data to select optimal tuning parameter based on averaged AUC. Under each scenario, we report results averaged over 100 simulated training datasets, each including a main-study component and an external-study component, and a single, large test dataset with a sample size of $10^6$. The predictive performance of models is evaluated using $\mathrm{R}^2$ for linear regression and AUC for logistic regression (Figure~\ref{fig:lgLasso}). We also report the uncertainty of model performance due to the randomness of training data by adding pointwise 95\% confidence bands in the figures.

We observe that for both linear and logistic regressions, the proposed HTL-GMM method with Lasso penalization leads to consistent improvement in the predictive performance of models over standard Lasso analysis of the main study across a wide range of sample sizes (Figure~\ref{fig:lgLasso}). As the sample size grows, the performance of both methods converges towards that of the true model indicating consistency of the Lasso-based model selection. In contrast, as the sample size increases, performance of the model built using external data alone quickly plateaus and stays far below the performance of the true model due to the exclusions of the true predictors in ${\bf w}$. For HTL-GMM, the use of variational GMM with the multiplicative-shrinkage (MS) or ridge kernel leads to significant improvement over the use of ordinary weighted GMM (owGMM) in a smaller sample size. Additional results shown in Supplemental Figure~\ref{fig:lgLasso1500}  demonstrate the importance of the use of variational GMM in even higher dimension ($p_{\mathrm{W}}=1500$) (Supplemental Figure~\ref{fig:lgLasso1500}). We observe the superiority of the MS-kernel over the ridge kernel in logistic regression task, while the opposite in linear regression task (Figure~\ref{fig:lgLasso}, Supplementary Figure~\ref{fig:lgLasso1500}, and \ref{fig:ps_ridge}). The observation is consistent with the literature regarding the tuning strategy for knowledge distillation \citep{hinton2015distilling, yim2017gift,clark2019bam,takamoto2020efficient}. 

\subsection{Post-selection Asymptotic Inference under Adaptive Lasso}

\begin{figure}[h!]
    \centering
    \includegraphics[width=.85\linewidth]{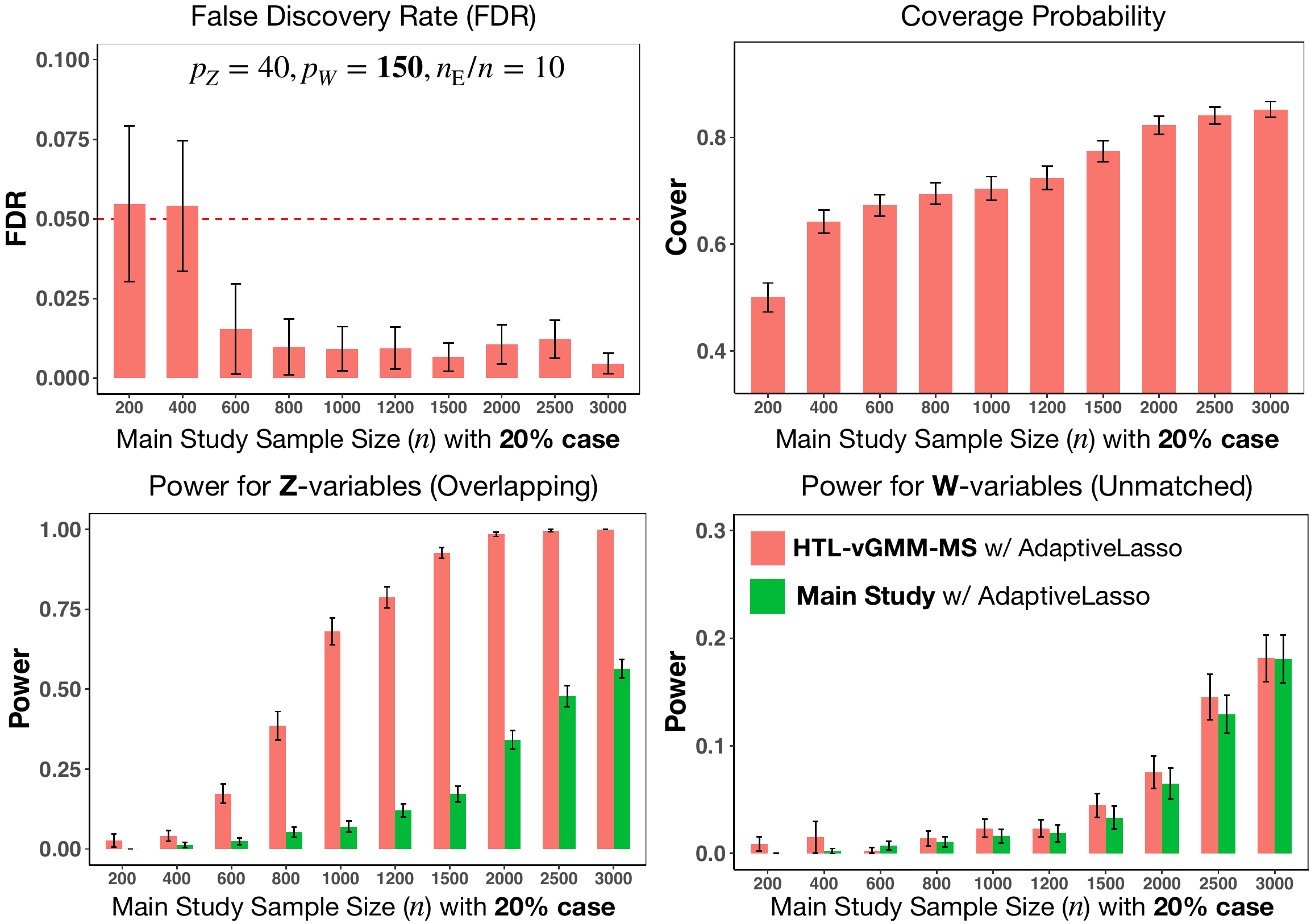}
    \caption{\textbf{Post-selection inference properties of HTL-GMM under adaptive Lasso and MS-kernel for logistic regression model.} Results are shown for standard adaptive Lasso analysis of the main study and HTL-GMM based integrated analysis of the main and external studies. The number of overlapping variables ($p_{\mathrm{Z}}$) is fixed at 40, the number of unmatched variables ($p_{\mathrm{W}}$) is 150, and the sample size of the external study relative to the main study ($n^{\mathrm{E}}/n$) is 10.}
    \label{fig:fdr}
\end{figure}

Next, we investigate the post-selection inference properties of HTL-GMM implemented with adaptive Lasso and variational GMM. Here, we consider the logistic regression setting with MS-kernel and $p_{\mathrm{Z}}=40,\,p_{\mathrm{W}}=150,\,n^{\mathrm{E}}/n = 10$. For each simulated dataset, we apply Benjamini-Hochberg (BH) multiple testing correction \citep{benjamini1995controlling} for the selected variables using associated adaptive Lasso $p$-values to maintain false discovery rate (FDR) at the 5\% level. We further evaluate the coverage probability for non-null variables based on the proportions of times the pointwise 95\% confidence intervals cover the true effect sizes, averaged over all the non-null variables. Results reported in Figure~\ref{fig:fdr} show that as the sample size increases, the method can maintain the desired FDR level. In a larger sample, the method appears to be conservative for FDR control, but this is likely due to the use of BH procedure and not the feature of the method itself. We further observe that the coverage of the pointwise confidence intervals continues to increase with sample sizes, but remains below nominal level even with fairly substantial sample size. Finally, we investigate the average power of HTL-GMM compared to the standard application of adaptive Lasso for selecting true non-null variables (Figure~\ref{fig:fdr}). We observe that, as expected, HTL-GMM gains major power for selecting non-null variables in ${\bf z}$ across different sample sizes. The method also shows some power gain when selecting non-null variables in ${\bf w}$ for smaller sample sizes. The results suggest that while there is no direct information on ${\bf w}$ in the external study, improvement of the power in selecting non-null ${\bf z}$ variables can lead to improvement of the power in selecting non-null ${\bf w}$ variables in the presence of correlations between variables across ${\bf w}$ and ${\bf z}$. However, such gain in power quickly disappears as the sample size increases. A similar trend is also observed when $p_{\mathrm{W}}$ is increased to 1,500, with all other conditions being the same (Supplementary Figure~\ref{fig:fdr1500}).

\section{A Real Data Application using UK Biobank}

In this section, we demonstrate an application of HTL-GMM for disease risk prediction using baseline risk factors and recently available proteomic data from the UK Biobank (UKB) study. The UK Biobank study is a large cohort involving about half a million individuals on whom baseline information on a large variety of health-related factors is collected \citep{sudlow2015uk}. Because of its large size, detailed data collection, ability to link to national population registries, and open data access, the study has become the bedrock of modern genetic and epidemiologic studies. We consider developing models for disease risk prediction using traditional risk factor data on all individuals and proteomic data released in March 2023 ($\#\mathrm{protein}= 1500$) on a subset of the study participants (50K). The data on 50K participants with proteomic data were recently used to develop risk prediction models for various diseases \citep{gadd2023blood}. We consider developing models using logistic regression for 10-year incidence risk for five common diseases: postmenopausal breast cancer (women only), colorectal cancer, cardiovascular disease (CVD), stroke, and asthma. We review the literature to identify known or suspected risk factors for each disease. Identified risk factors can be categorized into the following types: sociodemographics (e.g., age, sex), lifestyle and environmental factors (e.g., smoking status, alcohol consumption frequencies), physical measures (e.g., body mass index (BMI), height, low-density lipoprotein (LDL)), health and medical history, family history (e.g., mother or father's disease history), sex-specific factors (e.g., number of births for female participants), and polygenic risk scores released by UK Biobank \citep{thompson2022uk}. For the full list of risk factors, please refer to Table~\ref{tab:ukb} in the appendix.

We apply our proposed algorithm to combine data from unrelated white British ancestry individuals with complete data, i.e., those with risk factor and proteomic data (main study, $n \approx 30\mathrm{K}$) and those with only risk factor data (external study, $n^{\mathrm{E}} \approx 300\mathrm{K}$). For modeling the risk of incidence of a disease, we first remove individuals from the dataset who report prior history of the disease at the baseline. More details about data preprocessing can be found in supplemental materials. We report averaged AUC for each method for 5-fold nested cross-validation where an inner loop is use for optimal tuning parameter selection and the outer loop is used for evaluating  AUC in held-out test datasets. In Table~\ref{tab:1}, we present, for each disease, the number of cases in the main/external study, the number of underlying risk factors together with the cross-validated AUC of the proposed HTL-GMM with Lasso and standard Lasso when applied to data from the main study only. Given its superior performance observed in simulation studies for classification task, here we implement HTL-GMM with the variational MS-kernel. To examine the influence of the sample size of the main study, we also carry out an analysis only including a randomly selected subset of individuals from the main study with half the sample size.

We observe that across all diseases, HTL-vGMM-MS demonstrates consistent improvement in model performance. When we use the full sample size for the main study, the use of HTL-GMM leads to a 0.5\%-1\% increase in AUC value compared to standard Lasso across the diseases except for CVD. The result is intuitive as CVD is one of the most common diseases, and it includes a large number of cases in the main study itself.  When we use only half the sample in the main study, HTL-GMM leads to even more significant improvements compared to Lasso across the diseases. Overall, these results are consistent with those observed in simulation studies.  While the increase in model performance seems modest, they are not unremarkable because increasing AUC even by small percentage points can significantly impact the identification of individuals at the tail of the risk distribution. 
\par

\begin{table}
\centering
\setlength{\tabcolsep}{1pt}
\begin{tabular}{cccccccc}
\toprule Disease & $\begin{array}{c}\text {Postmenopausal} \\
\text { Breast Cancer} \end{array}$ & $\begin{array}{c}\text {Colorectal} \\
\text { Cancer}\end{array}$& \text {CVD} & \text {Stroke} & \text {Asthma} \\
\hline \# of Risk Factors  & 16 & 16  & 18  & 19  & 9  \\
\midrule
&\multicolumn{5}{c}{\textbf{With Full Sample Size for the Main Study}}                   \\
    \cmidrule(r){1-6}
$\begin{array}{c}\text {\# of Cases} \\
\text {(Main, External)}\end{array}$
& (520,5098)&(508,4619)&(2113,
17839)&(500,4210)&(501,4256)\\
\midrule
& \multicolumn{5}{c}{AUC (SD)}                   \\
    \cmidrule(r){2-6}
Main Study Lasso & 0.669\,(0.02)&0.698\,(0.02)	&0.775\,(0.01)&0.765\,(0.02)&0.675 \,(0.03)\\
 HTL-vGMM-MS Lasso & 0.681\,(0.02)&0.709\,(0.02)&0.775\,(0.01)&0.770\,(0.02)&0.682\,(0.03)\\
\bottomrule
&\multicolumn{5}{c}{\textbf{With Half the Sample Size for the Main Study}}              \\
    \cmidrule(r){1-6}
$\begin{array}{c}\text {\# of Cases} \\
\text {(Main, External)}\end{array}$
& (260,5098)&(254,4619)&(1057,
17839)&(250,4210)&(251,4256)\\
\midrule
& \multicolumn{5}{c}{AUC (SD)}                   \\
    \cmidrule(r){2-6}
Main Study Lasso & 0.659\,(0.03)&0.688\,(0.02)&0.770\,(0.01)&0.759\,(0.02)&0.657 \,(0.03)\\
 HTL-vGMM-MS Lasso &0.675\,(0.02)&0.706\,(0.02)&0.773\,(0.01)&0.768\,(0.02)&0.673\,(0.03)\\
\bottomrule
\end{tabular}
\caption{\textbf{Results from the UK Biobank data analysis on the risk prediction of five common diseases using traditional risk factors and proteomic data.} The whole cohort is divided into a ``main study'', consisting of individuals for whom both proteomics and risk factor data are available, and an ``external study'', consisting of individuals who have only risk factor data. Ten-year disease risks are modelled using logistic regression and the Lasso penalty for model regularization. For each disease, the number of incidence cases observed in the main and external studies and the number of risk factors included in the model are shown. Average values and standard deviations (SD) of AUC are reported for each method for 5-fold nested cross-validation, where a 10-fold inner loop is use for optimal tuning parameter selection and the outer loop is used for evaluating AUC in held-out test datasets. For each disease, models are built using standard Lasso analysis of the main study or the HTL-GMM with Lasso penalty based integrated analysis of the main and external studies. The HTL-GMM is implemented with the variational MS-kernel and initialized by the main-study Lasso estimator. The second half of the table shows when analysis are done with only half of the cases in the main-study.} 
    \label{tab:1}
\end{table}

\section{Discussion}

In this article, we have proposed and studied the properties of a method for building high-dimensional GLM models by combining individual-level data from a main study and information on a reduced model from an external study. Both our simulation studies and data analysis show the potential of the method for increasing predictive accuracy and efficiency of post-selection inference through the incorporation of external information. Our studies also provide a novel insight that, in high dimension, it is critical to consider variational GMM, which allows suitable regularization of the optimal GMM weight matrix. The proposed method can be extended in scenarios when there may be multiple external studies, each leading to a potentially unique ``reduced'' model depending on the features measured in the study. In this setting, a calibration equation can be set up for each external study based on its overlapping features with the main study. Then, a GMM-based penalized objective function can be defined to integrate information across all the different studies.  
\par 
We make an assumption of transportability for the reduced model parameters across studies after adjustment for respective design variables. While accounting for study-specific ``nuisance'' parameters can make the assumption more realistic, it still could be violated due to heterogeneity in the distribution of covariates across studies \citep{chatterjee2016constrained,kundu2019generalized,han2019empirical}. As the reduced model parameters are available from both main and external studies, a gross violation of the assumption could directly be checked by testing the equality of the two sets of parameters. Further, one could evaluate whether or not integrating information from external data improves model performance compared to the analysis of the main study alone. A more sophisticated approach to dealing with population heterogeneity would be to consider various shrinkage estimation approaches \citep{mukherjee2008exploiting,chen2009shrinkage,zhai2022data,gu2023meta}, which have been explored for trading-off bias versus efficiency for parameter estimation in low-dimensional settings. In our setting, a principled approach to accounting for population heterogeneity within the GMM framework will be to modify the estimating function ${\bf U}_2=0$ with ${\bf U}_2=\eta$, where $\eta$ is considered an unknown vector of parameters that would be estimated from the data itself, with incorporation of penalty so that it can be data-adaptively shrunken towards zero \citep{liao2013adaptive,cheng2015select}. 
\par
In our data application, we have access to individual-level data from both the ``main'' and ``external'' studies, which are both part of the same UKB cohort. In such a setting, an alternative approach for data integration would be to impute proteomic data on individuals not included in the main study. This could be achieved by training an imputation model using data from individuals in the main study. However, the development of models for imputing high-dimensional proteomic data, which has complex correlation structures among themselves and with other risk factors, can be daunting and will require strong modeling assumptions. Our setup  does not require any modeling of the joint distribution of $({\bf z}, {\bf w})$ because information is transferred from the external study through an outcome model. We have shown earlier that while misspecification of the outcome model for the external study does not introduce bias in parameter estimation of the target model, it can lead to some loss of efficiency \citep{kundu2023logistic}. Future research is merited in exploring alternative semi-parametric methods that do not require strong assumptions on the joint distribution of $({\bf z}, {\bf w})$ and yet can increase the efficiency of the analysis when individual-level data are available across different studies.  
\section*{Acknowledgement}
The work was supported by NIH grants R56HG013137 and R01HG010480. The UK Biobank data was accessed through application ID 17731.
\section*{Code Availability}
A software implementation, \texttt{htlgmm} for \texttt{R} \citep{Rcoreteam}, is available on GitHub at \url{https://github.com/RuzhangZhao/htlgmm}. Source code for simulation and data application, and a tutorial for how to use our package are available at \url{http://htlgmm.github.io/}.

\appendix
\section{Appendix}
\subsection{Algorithm}
\label{sec:comp}
\begin{algorithm*}[htbp]
\caption{HTL-GMM}\label{alg:HTL-GMM}
\hspace*{\algorithmicindent} \textbf{Input}: Main study: $({\bf y},{\bf X}=[{\bf A},{\bf Z},{\bf W}])$; Information from a pre-trained reduced model based on the external study: $(\tilde{\bb{\theta}}_{\mathrm{Z}}^{\mathrm{E}},\widetilde{\bf V}_{\bb{\theta}_{\mathrm{Z}}^{\mathrm{E}}})$.\\
 \hspace*{\algorithmicindent} \textbf{Output}: Estimated target coefficient $\hat{\bb{\beta}}_n$.
\begin{algorithmic}[1]
\State \textbf{Initialization}: Compute $\hat{\bb{\beta}}_{n}^{(0)}$ based on Lasso regression analysis of the main study with tuning parameter selected using 10-fold cross-validation; 
compute $\hat{\bb{\theta}}_{\mathrm{A}}$ by fitting the reduced model to the main study and define $\tilde{\bb{\theta}}^\top = [\hat{\bb{\theta}}_{\mathrm{A}}^\top,(\tilde{\bb{\theta}}_{\mathrm{Z}}^{\mathrm{E}})^\top]$.

\State \textbf{Weighting Matrix Estimation}: 
${\bf C}_n \leftarrow \left(\widehat{\mathrm{\bf V}}_{n}(\hat{\bb{\beta}}_{n}^{(0)},\tilde{\bb{\theta}},\hat{r})+\alpha_n\Big(\begin{array}{@{}c@{}@{}c@{}}
  {\bf K}_{11,n}\,\,   & {\bf 0} \\[-.5ex]
    {\bf 0}&{\bf 0}
\end{array}\Big)\right)^{-1}$ with selected kernel.

\State \textbf{Pseudo Matrix Transformation}: 
Compute ${\bf X}_{\mathrm{ps}} = {\bf X}_{\mathrm{ps}}(\hat{\bb{\beta}}_{n}^{(0)})$ and ${\bf y}_{\mathrm{ps}} = {\bf y}_{\mathrm{ps}}(\hat{\bb{\beta}}_{n}^{(0)},\tilde{\bb{\theta}})$ from (\ref{eq:ps:linear}) for linear and (\ref{eq:ps:logistic}) for logistic regressions.
\State \textbf{Target Coefficient}: Obtain $\hat{\bb{\beta}}_n\leftarrow\arg\min_{\bb{\beta}}\left\{\frac{1}{2}\bb{\beta}^\top{\bf X}_{\mathrm{ps}}^\top{\bf X}_{\mathrm{ps}}\bb{\beta} - \bb{\beta}^\top{\bf X}_{\mathrm{ps}}^\top{\bf y}_{\mathrm{ps}}+ \textbf{P}_{\lambda_n}(\bb{\beta})\right\}$ using the \texttt{glmnet} package for a fixed value of the tuning parameter and then obtain the optimal value of the tuning parameter through 10-fold cross-validation within the main study.

\State \textbf{Output}: $\hat{\bb{\beta}}_n$ for selected variables. Under adaptive Lasso, estimates of asymptotic standard errors are also returned.
\end{algorithmic}
\end{algorithm*}
\FloatBarrier

\subsection{Assumptions}
\begin{assumption*}[A0, Transportability] Refer to Section~\ref{sec:2.1}.
\end{assumption*}

\begin{assumption*}[A1, Behavior of initial estimator] $\hat{\bb{\beta}}_{n}^{(0)}\xrightarrow{p}\bb{\beta}^\star$ and has $\sqrt{n}$ convergence rate. 
\end{assumption*}

\begin{assumption*}[A2, Differentiability] 
${\bf U}_{l}({\bf x},y;\bb{\beta},\bb{\theta})$ is twice continuously differentiable for $\bb{\beta}$ and $\bb{\theta}$, where $(\bb{\beta},\bb{\theta})\in \mathcal{N}_c({\bb{\beta}}^\star)\times \mathcal{N}_c({\bb{\theta}}^\star)$, $\mathcal{N}_c({\bb{\beta}})$ and $\mathcal{N}_c({\bb{\theta}})$ are compact neighborhoods of $\bb{\beta}^\star$ and $\bb{\theta}^\star$, respectively, $l=1,2,3$.
\end{assumption*}
% \begin{assumption*}[AA] 
% \begin{equation*}
% \sup_{(\bb{\beta},\bb{\theta})\in B_{\bb{\beta}}\times B_{\bb{\theta}}}\left\| \frac{\partial^2 {\bf U}_{l,n}({\bf x},y;\bb{\beta},\bb{\theta})}{\partial \bb{\theta}^\top\partial \bb{\theta}}\right\|_2^2<+\infty, l=1,2,3.
% \end{equation*} 
% \end{assumption*}
\begin{assumption*}[A3, Full column rank for matrices]
$\mathbb{E}[\frac{\partial {\bf U}({\bf x},y;\bb{\beta},\bb{\theta})}{\partial \bb{\beta}}]$ and $\mathbb{E}[\frac{\partial {\bf U}_l({\bf x},y;\bb{\theta})}{\partial \bb{\theta}}],\, l=2,3$ are of full column rank. To match previous notations, we have $\bb{\Gamma}_{\bb{\beta}} = \mathbb{E}[\frac{\partial  {\bf U}({\bf x},y;\bb{\beta}^\star,\bb{\theta}^\star)}{\partial \bb{\beta}}]$, and $\bb{\Gamma}_{{\bf x}_{\mathrm{R}},{\bf x}_{\mathrm{R}}} = \mathbb{E}[\frac{\partial {\bf U}_3({\bf x}_{\mathrm{R}},y;\bb{\theta}^\star)}{\partial \bb{\theta}}]$.
\end{assumption*}

\begin{assumption*}[A4, Uniform convergence] 
\begin{align*}
&(i)\sup_{(\bb{\beta},\bb{\theta})\in \mathcal{N}_c({\bb{\beta}}^\star)\times \mathcal{N}_c({\bb{\theta}}^\star)} \| {\bf U}_{n}(\bb{\beta},\bb{\theta})- \mathbb{E}\left[{\bf U}({\bf x},y;\bb{\beta},\bb{\theta})\right]\|_2^2\xrightarrow{p} 0,\\
&(ii)\sup_{(\bb{\beta},\bb{\theta})\in \mathcal{N}_c({\bb{\beta}}^\star)\times \mathcal{N}_c({\bb{\theta}}^\star)} \left\| \frac{\partial {\bf U}_n(\bb{\beta},\bb{\theta})}{\partial \bb{\beta}} - \mathbb{E}\left[\frac{\partial {\bf U}({\bf x},y;\bb{\beta},\bb{\theta})}{\partial \bb{\beta}}\right]\right\|_2^2\xrightarrow{p} 0,\\
&(iii)\sup_{\bb{\theta}\in \mathcal{N}_c({\bb{\theta}}^\star)} \left\| \frac{\partial {\bf U}_{l,n}(\bb{\theta})}{\partial \bb{\theta}} - \mathbb{E}\left[\frac{\partial {\bf U}_l({\bf x},y;\bb{\theta})}{\partial \bb{\theta}}\right]\right\|_2^2\xrightarrow{p} 0, \,\,\, l=2,3.
\end{align*}

\end{assumption*}

\begin{assumption*}[A5]
${\bf C}_n$ is a symmetric, positive definite matrix, and $\|{\bf C}_n-{\bf C}\|_2^2\xrightarrow{p}0$, where ${\bf C}$ is finite, symmetric, and positive definite. 
\end{assumption*}
\begin{assumption*}[A6] $\mathrm{\bf V}(\bb{\beta},\bb{\theta},r)$ is of full rank, for $(\bb{\beta},\bb{\theta})\in \mathcal{N}_c({\bb{\beta}}^\star)\times \mathcal{N}_c({\bb{\theta}}^\star)$.  
\begin{equation*}
\left\|\widehat{\mathrm{\bf V}}_{n}(\hat{\bb{\beta}}_n^{(0)},\tilde{\bb{\theta}},\hat{r}) -\mathrm{\bf V}(\bb{\beta}^\star,\bb{\theta}^\star,r)\right\|_2^2\xrightarrow{p} 0,    
\end{equation*}
\begin{equation*}
\left\|\left(\widehat{\mathrm{\bf V}}_{n}(\hat{\bb{\beta}}_n^{(0)},\tilde{\bb{\theta}},\hat{r})+\alpha_n {\bf K}_n \right)^{-1} -\mathrm{\bf V}^{-1}(\bb{\beta}^\star,\bb{\theta}^\star,r)\right\|_2^2\xrightarrow{p} 0.    
\end{equation*}

\end{assumption*}
\subsection{Proofs}
\subsubsection{Proof of Lemma~\ref{lemma:U}}
\label{lemma:Uapp}

\begin{proof}
We expand $\sqrt{n}{\bf U}_n(\bb{\beta}^\star,\tilde{\bb{\theta}})$ with respect to $\bb{\theta}\in \mathcal{N}_c(\bb{\theta}^\star)$ by the first order Taylor's series expansion. Specifically, under Assumptions (A2) and (A4),
we derive the asymptotic distribution of ${\bf U}_n(\bb{\beta}^\star,\tilde{\bb{\theta}})$ as: 
\begin{align*}
\sqrt{n}{\bf U}_n(\bb{\beta}^\star,\tilde{\bb{\theta}})
&=\left(\begin{array}{cc}
     \sqrt{n}{\bf U}_{1,n}(\bb{\beta}^\star)
  \\
  \sqrt{n}{\bf U}_{2,n}(\bb{\beta}^\star,\bb{\theta}^\star) + \bb{\Gamma}_{{\bf z},{\bf x}_{\mathrm{R}}}\sqrt{n}(\tilde{\bb{\theta}} -\bb{\theta}^\star)
\end{array}\right)+o_{p}(1)\xrightarrow{d} \mathcal{N}\left({\bf 0},\left(\begin{array}{ccc}
  {\bf V}_{11}   & {\bf V}_{12} \\
    {\bf V}_{12}^\top & {\bf V}_{22}\\
\end{array}\right)\right).
\end{align*}
%where the Assumption (A2) guarantees the validity of the first order expansion, and the uniform convergence Assumption (A4(iii)) incorporates  $\bb{\Gamma}_{{\bf z},{\bf x}_{\mathrm{R}}}=\mathbb{E}\left[\frac{\partial {\bf U}_{2}({\bf x},y;\bb{\theta}^\star)}{\partial \bb{\theta}}\right]$. Under the uniform convergence for ${\bf U}_n$, i.e., Assumption (A4(i)), the consistency of $\tilde{\bb{\theta}}$ for $\bb{\theta}^\star$, we have the asymptotic mean to be ${\bf 0}$. \par 
Since $\tilde{\bb{\theta}}^\top = [\hat{\bb{\theta}}_{\mathrm{A}}^\top,(\tilde{\bb{\theta}}_{\mathrm{Z}}^{\mathrm{E}})^\top]$, $\bb{\Gamma}_{{\bf z},{\bf x}_{\mathrm{R}}}\sqrt{n}(\tilde{\bb{\theta}} -\bb{\theta}^\star)$ can be decomposed as $\bb{\Gamma}_{{\bf z},{\bf a}}\sqrt{n}(\hat{\bb{\theta}}_{\mathrm{A}} -\bb{\theta}^\star_{\mathrm{A}})+\bb{\Gamma}_{{\bf z},{\bf z}}\sqrt{n}(\tilde{\bb{\theta}}_{\mathrm{Z}}^{\mathrm{E}} -\bb{\theta}^\star_{\mathrm{Z}})$, involving estimates of parameters from main and external studies, respectively. Under Assumptions (A2)-(A4), we have $ \sqrt{n}(\hat{\bb{\theta}}_{\mathrm{A}}-\bb{\theta}_{\mathrm{A}}^{\star}) = \bb{\Gamma}^{{\bf a},{\bf x}_{\mathrm{R}}}\sqrt{n}{\bf U}_{3,n}(\bb{\theta}^\star)+o_p(1)$ and
\begin{equation*}
\sqrt{n}(\hat{\bb{\theta}}_{\mathrm{A}}-\bb{\theta}^{\star}_{\mathrm{A}})\xrightarrow{d} \mathcal{N}({\bf 0},{\bf V}_{\bb{\theta}_{\mathrm{A}}}) = \mathcal{N}\left({\bf 0},\bb{\Gamma}^{{\bf a},{\bf x}_{\mathrm{R}}}\mathbb{E}\left[    {\bf x}_{\mathrm{R}}{\bf x}_{\mathrm{R}}^\top 
\left\{\mu({\bf x}_{\mathrm{R}}^\top\bb{\theta}^\star) - {\bf y}\right\}^2\right]\bb{\Gamma}^{{\bf x}_{\mathrm{R}},{\bf a}}\right),
\label{eq:sup:hatthetaa}
\end{equation*}
where $\bb{\Gamma}_{{\bf x}_{\mathrm{R}},{\bf x}_{\mathrm{R}}} = \mathbb{E}[ \frac{\partial{{\bf U}_3({\bf x}_{\mathrm{R}},y;\bb{\theta}^\star)}}{\partial \bb{\theta}}]$, and $\bb{\Gamma}_{ {\bf x}_{\mathrm{R}},{\bf x}_{\mathrm{R}}}^{-1} = [\bb{\Gamma}^{ {\bf x}_{\mathrm{R}},{\bf a}},\bb{\Gamma}^{ {\bf x}_{\mathrm{R}},{\bf z}}] = [(\bb{\Gamma}^{{\bf a},{\bf x}_{\mathrm{R}}})^\top,(\bb{\Gamma}^{{\bf z},{\bf x}_{\mathrm{R}}})^\top]$. Next, for $\tilde{\bb{\theta}}_{\mathrm{Z}}^{\mathrm{E}}$ from external study, we have
\begin{equation*}
\sqrt{n}(\tilde{\bb{\theta}}_{\mathrm{Z}}^{\mathrm{E}} -\bb{\theta}^\star_{\mathrm{Z}}) = \sqrt{\frac{n}{n^{\mathrm{E}}}}\sqrt{n^{\mathrm{E}}}(\tilde{\bb{\theta}}_{\mathrm{Z}}^{\mathrm{E}} -\bb{\theta}^\star_{\mathrm{Z}})\xrightarrow{d}  \mathcal{N}\left({\bf 0},r^{-1}{\bf V}_{\bb{\theta}_{\mathrm{Z}}^{\mathrm{E}}}\right),
\label{eq:sup:tildethetaz}
\end{equation*}
since $\lim_{n\rightarrow +\infty} n^{\mathrm{E}}/n  = r, 0<r<\infty$. Because $\tilde{\bb{\theta}}_{\mathrm{Z}}^{\mathrm{E}} $ comes from external study, which is independent of main study, its asymptotic covariance with all the other terms is ${\bf 0}$. 

Combing all of the above and using the central limit theorem and Slutsky's theorem, we have the asymptotic variance ${\bf V}_{11} = {\bf V}_{{\bf x},{\bf x}} = \mathbb{E}\left[{\bf U}_1({\bf x},y;\bb{\beta}^\star){\bf U}_1({\bf x},y;\bb{\beta}^\star)^\top\right]$,
\begin{align*}
&{\bf V}_{22} = \text{Var}\left\{  {\bf U}_2({\bf x};\bb{\beta}^\star,\bb{\theta}^\star)+\bb{\Gamma}_{{\bf z},{\bf a}}\bb{\Gamma}^{{\bf a},{\bf x}_{\mathrm{R}}}{\bf U}_{3}({\bf x}_{\mathrm{R}},y;\bb{\theta}^\star) \right\} + \bb{\Gamma}_{{\bf z},{\bf z}}\left(r^{-1}{\bf V}_{\bb{\theta}_{\mathrm{Z}}^{\mathrm{E}}}\right) \bb{\Gamma}_{{\bf z},{\bf z}} \\ 
&= {\bf V}_{{\bf z},{\bf z}} + \bb{\Gamma}_{{\bf z},{\bf a}}{\bf V}_{\bb{\theta}_{\mathrm{A}}} \bb{\Gamma}_{{\bf a},{\bf z}}+ {\bf V}_{{\bf z},{\bf x}_{\mathrm{R}}}\bb{\Gamma}^{{\bf x}_{\mathrm{R}},{\bf a}}\bb{\Gamma}_{{\bf a},{\bf z}} + ({\bf V}_{{\bf z},{\bf x}_{\mathrm{R}}}\bb{\Gamma}^{{\bf x}_{\mathrm{R}},{\bf a}}\bb{\Gamma}_{{\bf a},{\bf z}})^\top+\bb{\Gamma}_{{\bf z},{\bf z}}\left(r^{-1}{\bf V}_{\bb{\theta}_{\mathrm{Z}}^{\mathrm{E}}}\right) \bb{\Gamma}_{{\bf z},{\bf z}},
\end{align*}
where ${\bf V}_{{\bf z},{\bf z}} =\mathbb{E}\left[{\bf U}_2({\bf x};\bb{\beta}^\star,\bb{\theta}^\star){\bf U}_2({\bf x};\bb{\beta}^\star,\bb{\theta}^\star)^\top\right]$, ${\bf V}_{{\bf z},{\bf x}_{\mathrm{R}}} = \mathbb{E}\left[{\bf U}_2({\bf x};\bb{\beta}^\star,\bb{\theta}^\star){\bf U}_3({\bf x}_{\mathrm{R}},y;\bb{\theta}^\star)^\top\right]$, and asymptotic covariance, 
\begin{align*}
{\bf V}_{12} = \text{Cov}\left\{ {\bf U}_1({\bf x},y;\bb{\beta}^\star),{\bf U}_2({\bf x};\bb{\beta}^\star,\bb{\theta}^\star)+\bb{\Gamma}_{{\bf z},{\bf a}}\bb{\Gamma}^{{\bf a},{\bf x}_{\mathrm{R}}}{\bf U}_{3}({\bf x}_{\mathrm{R}},y;\bb{\theta}^\star)\right\} = {\bf V}_{{\bf x},{\bf z}}+{\bf V}_{{\bf x},{\bf x}_{\mathrm{R}}}\bb{\Gamma}^{{\bf x}_{\mathrm{R}},{\bf a}}\bb{\Gamma}_{{\bf a},{\bf z}},
\end{align*}
where ${\bf V}_{{\bf x},{\bf z}} = \mathbb{E}\left[{\bf U}_1({\bf x},y;\bb{\beta}^\star){\bf U}_2({\bf x};\bb{\beta}^\star,\bb{\theta}^\star)^\top\right]$, ${\bf V}_{{\bf x},{\bf x}_{\mathrm{R}}}= \mathbb{E}\left[{\bf U}_1({\bf x},y;\bb{\beta}^\star){\bf U}_3({\bf x}_{\mathrm{R}},y;\bb{\theta}^\star)^\top\right]$.
\end{proof}
\subsubsection{Proof of Theorem~\ref{thm:oracle2}}
\label{sec:proof:thm1}

\begin{proof}

The proof mainly follows the techniques used in Theorem~2 of \cite{zou2006adaptive} and Theorem~5 of \cite{zou2008one}. Let $\bb{\beta} = \bb{\beta}^\star + \frac{{\bf u}}{\sqrt{n}}$, and 
\begin{equation*}
\Psi_n({\bf u}) = \left\|{\bf y}_{\mathrm{ps}}^{(0)} - \sum_{j=1}^{p_{\mathrm{X}}}{\bf x}_{\mathrm{ps},j}^{(0)}(\beta_j^\star + \frac{u_j}{\sqrt{n}}) \right\|_2^2 + \lambda_n \sum_{j=1}^{p_{\mathrm{X}}} \hat{w}_j|\beta_j^\star + \frac{u_j}{\sqrt{n}}|,
\end{equation*}
where ${\bf y}_{\mathrm{ps}}^{(0)}={\bf y}_{\mathrm{ps}}(\hat{\bb{\beta}}_n^{(0)},\tilde{\bb{\theta}}) = \sqrt{n}{\bf C}_n^{1/2}\left\{\frac{\partial {\bf U}_n(\hat{\bb{\beta}}_n^{(0)},\tilde{\bb{\theta}})}{\partial \bb{\beta}}\hat{\bb{\beta}}_n^{(0)} - {\bf U}_n(\hat{\bb{\beta}}_n^{(0)},\tilde{\bb{\theta}})\right\}$, ${\bf X}_{\mathrm{ps}}^{(0)} ={\bf X}_{\mathrm{ps}} (\hat{\bb{\beta}}_n^{(0)},\tilde{\bb{\theta}})= \sqrt{n}{\bf C}_n^{1/2}\frac{\partial {\bf U}_n(\hat{\bb{\beta}}_n^{(0)},\tilde{\bb{\theta}})}{\partial \bb{\beta}}$ and ${\bf x}_{\mathrm{ps},j}^{(0)}$ is the $j$-th column ${\bf X}_{\mathrm{ps}}^{(0)}$. Let $\hat{\bf u}_n = \arg\min \Psi_n({\bf u})$ and we have $\hat{\bb{\beta}}^{\mathrm{aLasso},(n)} = \bb{\beta}^\star + \frac{\hat{\bf u}_n}{\sqrt{n}}$. Following the proof of Theorem~2 of \cite{zou2006adaptive}, we consider 
\begin{align*}
&\Psi_n({\bf u}) - \Psi_n({\bf 0}) = {\bf u}^\top\frac{{{\bf X}_{\mathrm{ps}}^{(0)}}^\top{\bf X}_{\mathrm{ps}}^{(0)}}{n}{\bf u} - 2\frac{{\bb{\epsilon}_{\mathrm{ps}}^{(0)}}^\top{\bf X}_{\mathrm{ps}}^{(0)}}{\sqrt{n}}{\bf u}+\lambda_n \sum_{j=1}^{p_{\mathrm{X}}} \hat{w}_j\left(|\beta_j^\star + \frac{u_j}{\sqrt{n}}|-|\beta_j^\star|\right),
\end{align*}
where $\bb{\epsilon}_{\mathrm{ps}}^{(0)} = {\bf y}_{\mathrm{ps}}^{(0)} - {\bf X}_{\mathrm{ps}}^{(0)}\bb{\beta}^\star$, and we need to consider the asymptotic behavior of $\frac{1}{n}{{\bf X}_{\mathrm{ps}}^{(0)}}^\top{\bf X}_{\mathrm{ps}}^{(0)}$ and $\frac{1}{\sqrt{n}}{\bb{\epsilon}_{\mathrm{ps}}^{(0)}}^\top{\bf X}_{\mathrm{ps}}^{(0)}$. Under Assumptions (A1)-(A2), we can expand ${\bf U}_n(\bb{\beta}^\star,\tilde{\bb{\theta}}) = {\bf U}_n(\hat{\bb{\beta}}_n^{(0)},\tilde{\bb{\theta}}) + \frac{\partial {\bf U}_n(\hat{\bb{\beta}}_n^{(0)},\tilde{\bb{\theta}})}{\partial \bb{\beta}}(\bb{\beta}^\star - \hat{\bb{\beta}}_n^{(0)}) + o_p(\frac{1}{\sqrt{n}})$. So, we have $\bb{\epsilon}_{\mathrm{ps}}^{(0)} = {\bf y}_{\mathrm{ps}}^{(0)} - {\bf X}_{\mathrm{ps}}^{(0)}\bb{\beta}^\star = \sqrt{n}{\bf C}_n^{1/2}\Big\{\frac{\partial {\bf U}_n(\hat{\bb{\beta}}_n^{(0)},\tilde{\bb{\theta}})}{\partial \bb{\beta}}(\hat{\bb{\beta}}_n^{(0)} - \bb{\beta}^\star)- {\bf U}_n(\bb{\beta}^\star,\tilde{\bb{\theta}})$ $ -\frac{\partial {\bf U}_n(\hat{\bb{\beta}}_n^{(0)},\tilde{\bb{\theta}})}{\partial \bb{\beta}}(\hat{\bb{\beta}}_n^{(0)} - \bb{\beta}^\star) \Big\}+o_p(1)=- \sqrt{n}{\bf C}_n^{1/2}{\bf U}_n(\bb{\beta}^\star,\tilde{\bb{\theta}})+o_p(1)$. Under Assumptions (A1)-(A5) and Slutsky's theorem, we have $\frac{1}{n}{{\bf X}_{\mathrm{ps}}^{(0)}}^\top{\bf X}_{\mathrm{ps}}^{(0)} = \frac{\partial {\bf U}_n(\hat{\bb{\beta}}_n^{(0)},\tilde{\bb{\theta}})^\top}{\partial \bb{\beta}} {\bf C}_n \frac{\partial {\bf U}_n(\hat{\bb{\beta}}_n^{(0)},\tilde{\bb{\theta}})}{\partial \bb{\beta}} \xrightarrow{p}\bb{\Gamma}_{\bb{\beta}}^\top{\bf C}\bb{\Gamma}_{\bb{\beta}}$. Additionally, under Lemma~\ref{lemma:U}, $\left(\frac{1}{\sqrt{n}}{\bb{\epsilon}_{\mathrm{ps}}^{(0)}}^\top{\bf X}_{\mathrm{ps}}^{(0)}\right) \xrightarrow{d} \mathcal{N}({\bf 0},\bb{\Gamma}_{\bb{\beta}}^\top{\bf C}\mathrm{\bf V}{\bf C}\bb{\Gamma}_{\bb{\beta}})$.

Moreover, the limiting behavior of $\lambda_{n}$ meets the requirement for adaptive Lasso with linear regression setup, detailed in Theorem~2 of \cite{zou2006adaptive}. We, therefore, can derive (1) the consistency in variable selection, and (2) the asymptotic normality: $\sqrt{n}\left(  
\hat{\boldsymbol{\beta}}^{\mathrm{aLasso},(n)}_{\mathcal{A}^\star} - \bb{\beta}^\star_{\mathcal{A}^\star}\right) \xrightarrow{d} \mathcal{N}\left({\bf 0}, \boldsymbol{\Sigma}^\star_{\mathcal{A}^\star}  \right)$. Following \cite{zou2006adaptive}, we know the asymptotic variance for parameters with the true nonzero effects is the asymptotic variance of $\left\{(\frac{1}{n}{{\bf X}_{\mathrm{ps}}^{(0)}}^\top{\bf X}_{\mathrm{ps}}^{(0)})_{\mathcal{A}^\star}\right\}^{-1}\left\{ (
\frac{1}{\sqrt{n}}{{\bf X}_{\mathrm{ps}}^{(0)}}^\top\bb{\epsilon}_{\mathrm{ps}}^{(0)})_{\mathcal{A}^\star}\right\}$. So, $\boldsymbol{\Sigma}^\star_{\mathcal{A}^\star}= \{(\bb{\Gamma}_{\bb{\beta}}^\top{\bf C}\bb{\Gamma}_{\bb{\beta}})_{\mathcal{A}^\star}\}^{-1}\{( \bb{\Gamma}_{\bb{\beta}}^\top{\bf C}\mathrm{\bf V}{\bf C}\bb{\Gamma}_{\bb{\beta}} )_{\mathcal{A}^\star}\} \{ ( \bb{\Gamma}_{\bb{\beta}}^\top{\bf C}\bb{\Gamma}_{\bb{\beta}} )_{\mathcal{A}^\star}\}^{-1}$. Moreover, under Assumption (A6), when ${\bf C}_n =\left(\widehat{\mathrm{\bf V}}_{\mathrm{opt},n}+\alpha_n{\bf K}_n\right)^{-1}$, ${\bf C}_n\xrightarrow{p} \mathrm{\bf V}^{-1}$, and we achieve (3) the optimality for $\bb{\Sigma}^\star_{\mathcal{A}^\star}= \{(\bb{\Gamma}_{\bb{\beta}}^\top\mathrm{\bf V}^{-1}\bb{\Gamma}_{\bb{\beta}})_{\mathcal{A}^\star}\}^{-1}$ by Slutsky's theorem under the condition $\alpha_n = o(1)$. 
\end{proof}

\newpage

\putbib
\end{bibunit}
\begin{bibunit}
    
\newpage
\section*{Supplementary Material}
\beginsupplement
\setcounter{section}{0}
\section{Supplementary Figures}
\label{sec:fig:predict}
\renewcommand{\thefigure}{S1}
\begin{figure}[h!]
    \centering
    \includegraphics[width=.8\linewidth]{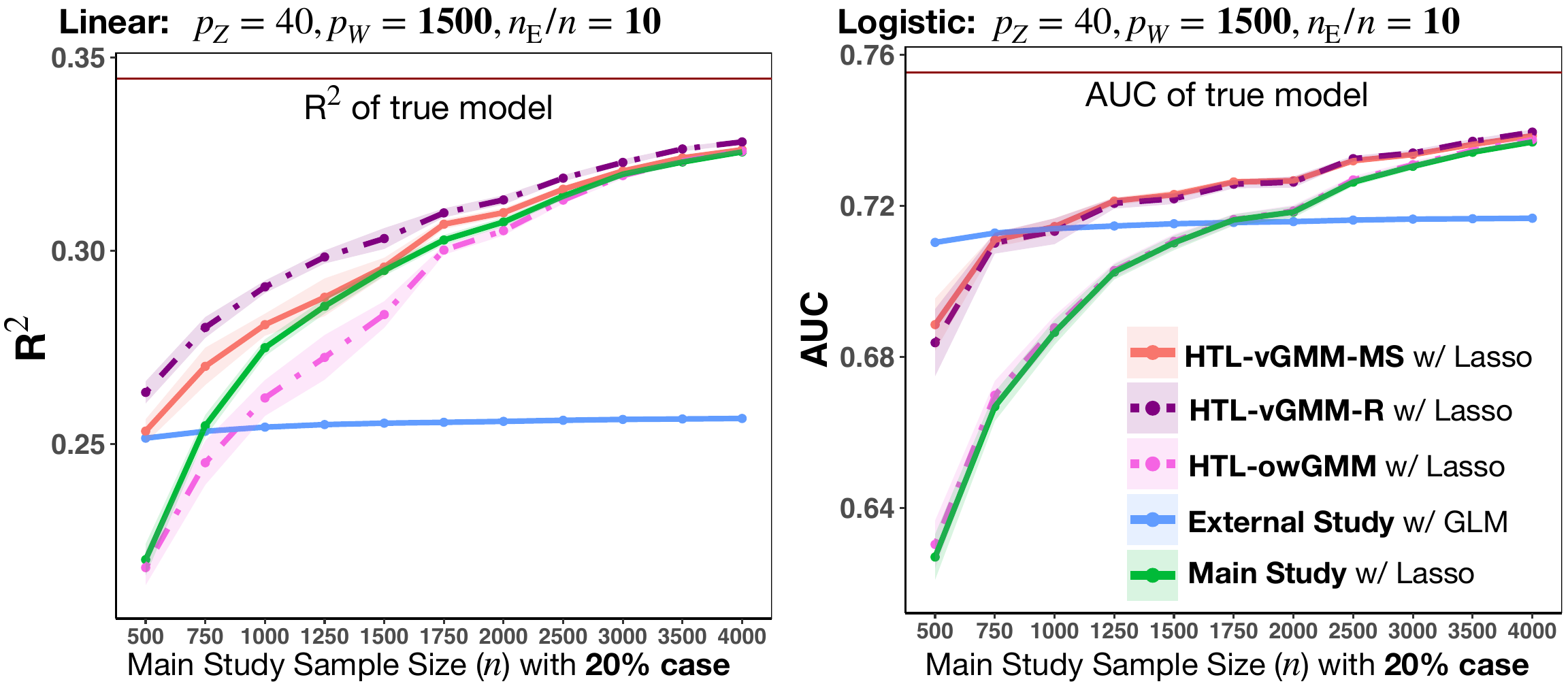}
    \caption{\textbf{Simulation study results showing the predictive performance of HTL-GMM and alternative methods with Lasso penalization for linear and logistic regression models with $\bb{p_{\mathrm{W}}}$ changed to 1,500.} The methods included are HTL-GMM-MS which uses multiplicative-shrinkage variational kernel, HTL-GMM-R which uses ridge variational kernel, HTL-owGMM which uses ordinary GMM, standard analysis of only the main study or external study in the linear or logistic regression setting. For HTL-GMM and standard analysis of the main study, the full models are fitted using the Lasso penalty function. The reduced model for the external study is fitted using standard linear or logistic regression. The prediction performance of models, quantified by $\mathrm{R}^2$ or AUC, is evaluated based on a large validation sample simulated independent of the main and external studies, while the prediction performance of true model is marked by top red lines. The sample size of the main study varies along the $x$-axis for each figure. Compared with Figure~\ref{fig:lgLasso}, the number of unmatched variables ($p_{\mathrm{W}}$) changes from 150 to 1,500, while the number of overlapping variables ($p_{\mathrm{Z}}$) is fixed at 40, and the sample size of the external study relative to the main study ($n^{\mathrm{E}}/n$) is 10.}
    \label{fig:lgLasso1500}
\end{figure}

\renewcommand{\thefigure}{S2}
\begin{figure}[H]
    \centering
\includegraphics[width=.8\linewidth]{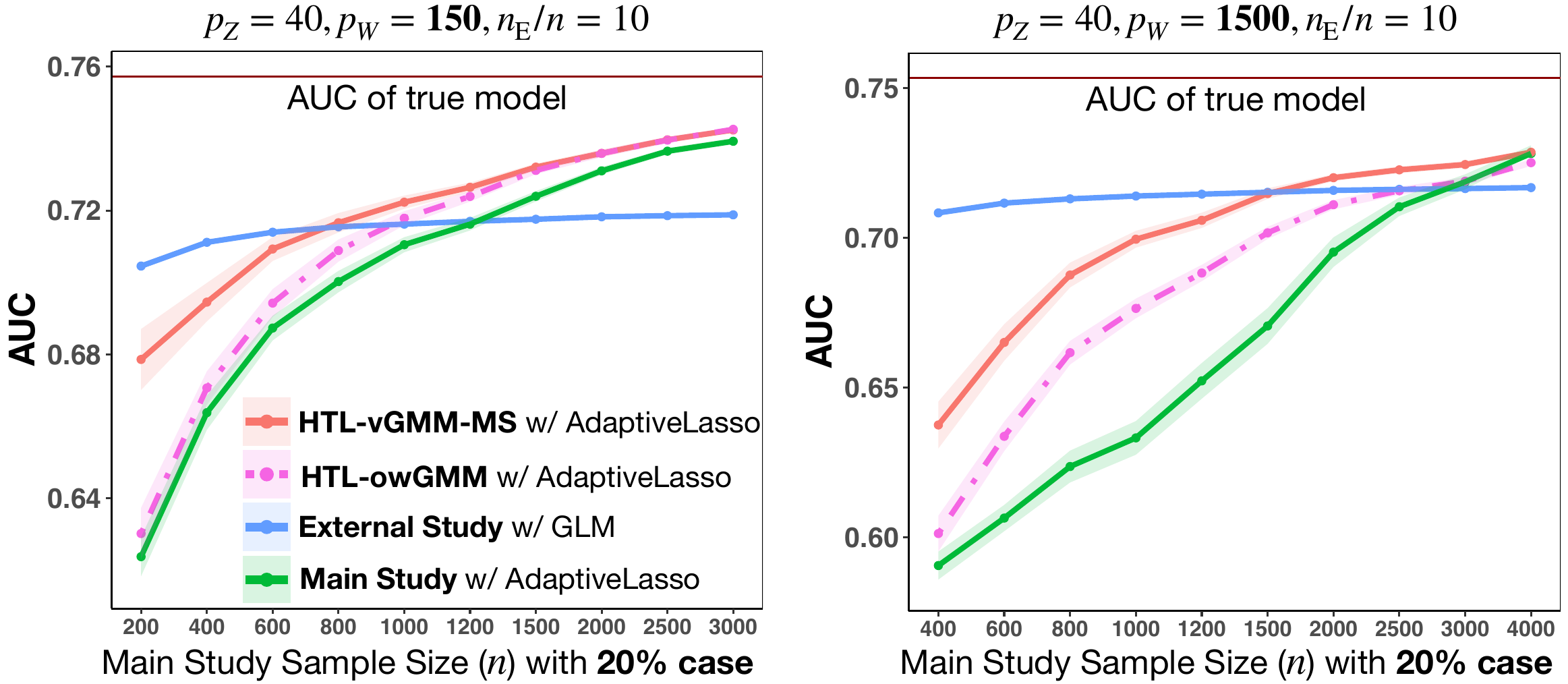}
    \caption{\textbf{Simulation study results showing the predictive performance of HTL-GMM and alternative methods with adaptive Lasso penalization for logistic regression model.} The methods included are HTL-GMM-MS which uses multiplicative-shrinkage variational kernel, HTL-owGMM which uses ordinary GMM, standard analysis of only the main study or external study in the logistic regression setting. For HTL-GMM and standard analysis of the main study, the full models are fitted using the adaptive Lasso penalty function. The reduced model for the external study is fitted using standard logistic regression. The prediction performance of models, quantified by AUC, is evaluated based on a large validation sample simulated independent of the main and external studies, while the prediction performance of true model is marked by top red lines. The sample size of the main study varies along the $x$-axis for each figure. The number of unmatched variables ($p_{\mathrm{W}}$) ranges from 150 to 1,500, while the number of overlapping variables ($p_{\mathrm{Z}}$) is fixed at 40, and the sample size of the external study relative to the main study ($n^{\mathrm{E}}/n$) is 10.}
    \label{fig:lgada}
\end{figure}
\renewcommand{\thefigure}{S3}
\begin{figure}[h!]
    \centering
    \includegraphics[width=.85\linewidth]{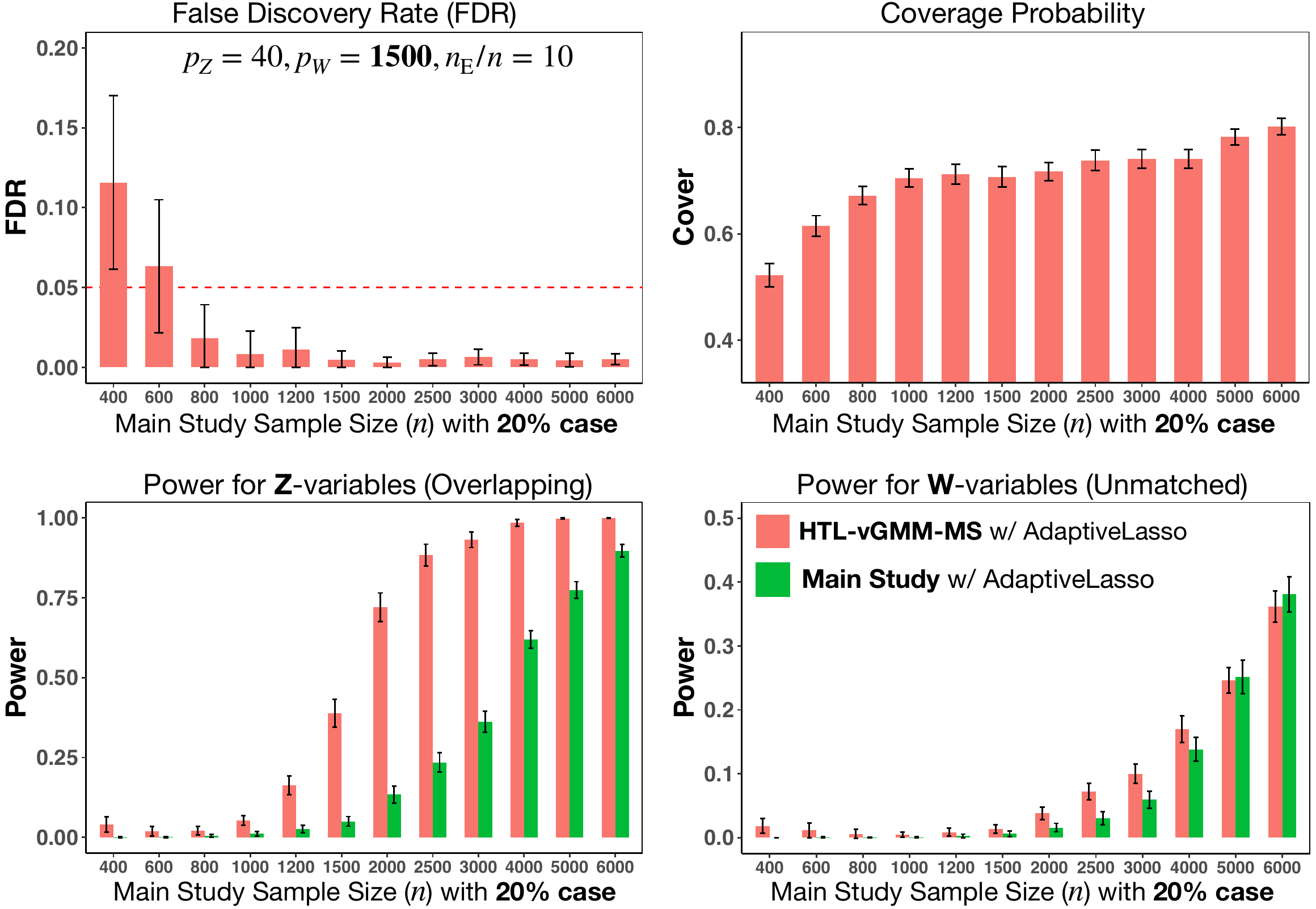}
    \caption{\textbf{Post-selection inference properties of HTL-GMM under adaptive Lasso and MS-kernel for logistic regression model with $\bb{p_{\mathrm{W}}}$ changed to 1,500.} Results are shown for standard adaptive Lasso analysis of the main study and HTL-vGMM-MS based integrated analysis of the main and external studies. The number of overlapping variables ($p_{\mathrm{Z}}$) is fixed at 40, the number of unmatched variables ($p_{\mathrm{W}}$) is 1,500, and the sample size of the external study relative to the main study ($n^{\mathrm{E}}/n$) is 10.}
    \label{fig:fdr1500}
\end{figure}

\renewcommand{\thefigure}{S4}
\begin{figure}[h!]
    \centering
    \includegraphics[width=.85\linewidth]{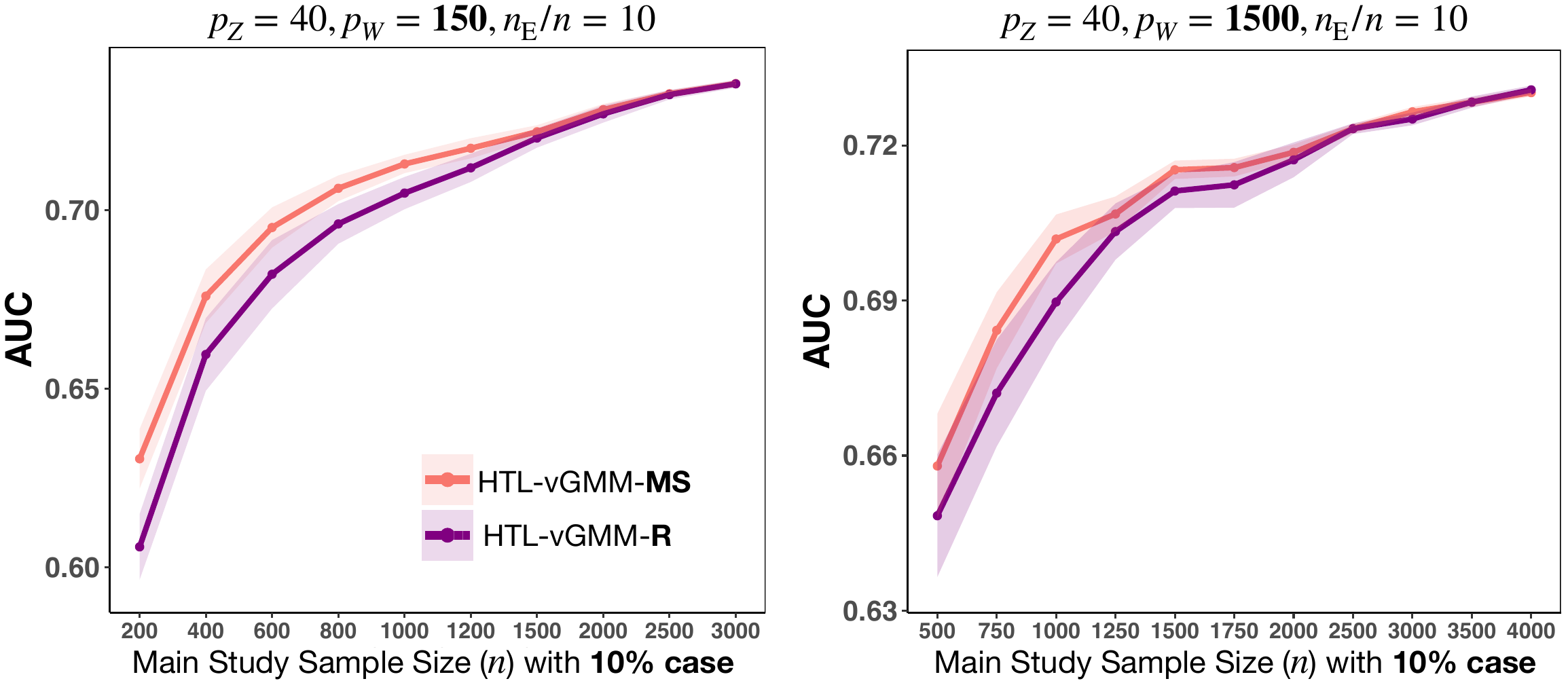}
    \caption{\textbf{Simulation study results showing the predictive performance of HTL-GMM for logistic regression models with a case ratio 10\%.} The methods included are HTL-GMM-MS which uses multiplicative-shrinkage variational kernel, HTL-GMM-R which uses ridge variational kernel for comparison. The prediction performance of models, quantified by AUC, is evaluated based on a large validation sample simulated independent of the main and external studies. The sample size of the main study varies along the $x$-axis for each figure. The number of unmatched variables ($p_{\mathrm{W}}$) ranges from 150 to 1,500, while the number of overlapping variables ($p_{\mathrm{Z}}$) is fixed at 40, and the sample size of the external study relative to the main study ($n^{\mathrm{E}}/n$) is 10.}
    \label{fig:ps_ridge}
\end{figure}

\FloatBarrier
\section{Supplementary Notes for UK Biobank Risk Factors}
For preprocessing UKB proteomics data, we adhere to the steps outlined by \cite{gadd2023blood}. For UKB risk factor data, we use the data collected on each participant's entry date and mark any absence of data collection on entry date as missing. For risk factors including smoking status, alcohol frequencies, we treat them as categorical variables. 

We apply the following five steps of filtering. First, we exclude participants who express a desire to withdraw from the study. Second, we exclude the related individuals based on relationships among participants provided by the UK Biobank study. Third, we exclude participants with discrepancies between reported and genetic sex. Fourth, restrict analysis to individuals who self-identify as white ancestry. Fifth, we only include those risk factors that have a moderate to small amount of missing data. In particular, we exclude any continuous variable with more than a 5\% missing rate and any categorical variable with more than a 1\% missing rate. For continuous risk factors included in the model, we replace missing values with observed mean values. Finally, we remove individuals if they have missing values for any of the categorical risk factors.

Following these data filtering stages, the preliminary sample sizes for the underlying cohorts are approximately 140K for breast cancer, 375K for colorectal cancer, 360K for CVD, 376K for stroke, and 348K for asthma. Furthermore, we integrate the preprocessed proteomics data with the corresponding risk factor data by participant ID to establish the main study. The external study consists of the participants with risk factor data but without aligned proteomics data. The sample sizes for main/external studies are approximately 14K/126K for breast cancer, 37K/338K for colorectal cancer, 35K/325K for CVD, 37K/339K for stroke, and 34K/313K for asthma. 
\par
Based on the literature, we compile a list of known or suspected risk factors for each disease, which are available in UKB data, and detailed on their website \citep{sudlow2015uk}. The referenced literature includes breast cancer \citep{sun2017risk}, colorectal cancer \citep{rawla2019epidemiology}, CVD \citep{damen2016prediction}, stroke \citep{boehme2017stroke},  asthma \citep{beasley2015risk}. For each disease, we also add polygenic risk score \citep{zhang2023new,zhang2024ensemble} released by UK Biobank \citep{thompson2022uk} as genetic risk factor. All the risk factors we summarize from the literature are concluded in Table~\ref{tab:ukb}.

\renewcommand{\thetable}{S1}
\begin{sidewaystable}[ph!]
    \centering
\begin{adjustbox}{width=20cm,center}   
\setlength{\tabcolsep}{1pt}
    \begin{tabular}{|c|c|c|c|c|c|c|c|c|c|c|c|c|c|c|c|c|c|c|c|c|}
\hline \begin{tabular}{cc} 
\rotatebox[origin=c]{45}{
\scalebox{2}[2]{$\backslash$}}&Risk factor\\
Disease&\rotatebox[origin=c]{45}{\scalebox{2}[2]{$\backslash$}} \end{tabular}  & Demographics & \begin{tabular}{c}Lifestyle and\\ Environment \end{tabular} & \begin{tabular}{c}Physical \\Measures \end{tabular} & \begin{tabular}{c}Health and\\ Medical History \end{tabular} & \begin{tabular}{c}Family \\ History \end{tabular} & \begin{tabular}{c}Sex-\\specific\\ Factors \end{tabular} \\
\hline \begin{tabular}{c} Breast  \\
 Cancer \end{tabular} & age. &\begin{tabular}{c} alcohol frequencies,\\ smoking status, \\ smoking pack years. \end{tabular} & 
height, BMI.&
\begin{tabular}{c}ever had hormone \\ replacement therapy, \\
ever used oral \\contraceptive, \\ history of benign\\ breast disease. \end{tabular}
&\begin{tabular}{c}breast cancer \\ history of \\ mother, siblings.\end{tabular}&
\begin{tabular}{c}
age at menarche,\\ number of birth, \\ ever had still \\ birth or termination,\\ ever had bilateral \\ oophorectomy.
\end{tabular}\\
\hline \begin{tabular}{c} Colorectal  \\
 Cancer \end{tabular} &age, sex.&\begin{tabular}{c} alcohol frequencies, \\ smoking status, \\ smoking pack years,\\ physical activities, \\ processed meat.\end{tabular}&height, BMI.&\begin{tabular}{c} history of polyps\\history of type II diabetes,\\ history of Crohn’s disease.
\end{tabular}&\begin{tabular}{c} bowel cancer \\ history of \\ father, mother, \\siblings.
\end{tabular}&  \\
\hline  CVD  & age, sex.&\begin{tabular}{c}  alcohol frequencies, \\ smoking status,\\ smoking pack years.\end{tabular}&\begin{tabular}{c} systolic blood pressure,\\ BMI, pulse rate, LDL, \\ C-reactive protein (crp),\\ triglyceride.\end{tabular}&\begin{tabular}{c} history of type II diabetes,\\ history of hypertension,\\ history of atrial fibrillation.\end{tabular}&\begin{tabular}{c} heart disease \\ history of \\ father, mother, \\ siblings.\end{tabular} & \\
\hline  Stroke  & age, sex. &\begin{tabular}{c} alcohol frequencies,\\ smoking status,\\smoking pack years,\\physical activities.\end{tabular}& \begin{tabular}{c}
systolic blood pressure, \\ BMI, LDL, \\C-reactive protein (crp),\\ triglyceride.\end{tabular}& \begin{tabular}{c}history of type II diabetes,\\ history of hypertension,\\ history of atrial fibrillation, \\ history of CVD.\end{tabular} & \begin{tabular}{c}stroke history of \\
 father, mother,\\ siblings.\end{tabular} & \\
\hline  Asthma  & age, sex. &\begin{tabular}{c}
 alcohol frequencies,\\smoking status,\\ smoking pack years.\end{tabular} &BMI.&\begin{tabular}{c}PM2.5 level, \\
whether mother \\smokes around birth.\end{tabular} & &  \\
\hline
\end{tabular}
\end{adjustbox}
    \caption{Summary of risk factor data from the UK Biobank to support the data analysis on the risk prediction of five common diseases.}
    \label{tab:ukb}
\end{sidewaystable}
\putbib
\end{bibunit}
\end{document}